\documentclass{article}

\usepackage{graphicx}
\usepackage[margin=1.4in]{geometry}
\usepackage{amsfonts}
\usepackage{amsmath}
\usepackage[english]{babel}
\usepackage{hyperref}
\usepackage{color}
\hypersetup{colorlinks=true,citecolor=blue,linkcolor=blue,filecolor=blue,urlcolor=blue}

\newcommand{\bv}{\begin{verse}}
\newcommand{\ev}{\end{verse}}
\newcommand{\be}{\begin{equation}}
\newcommand{\ee}{\end{equation}}
\newcommand{\bea}{\begin{eqnarray}}
\newcommand{\eea}{\end{eqnarray}}
\newcommand{\bq}{\begin{quotation}}
\newcommand{\eq}{\end{quotation}}

\newcommand{\Vaxjo}{V\"axj\"o}

\newtheorem{mohrhoff}{Ulrichism}
\newcommand{\bum}{\begin{mohrhoff}\protect$\!\!${\em\bf :}$\;\;$}
\newcommand{\eum}{\end{mohrhoff}}

\newtheorem{gefter}{Gefterism}
\newcommand{\bag}{\begin{gefter}\protect$\!\!${\em\bf :}$\;\;$}
\newcommand{\eag}{\end{gefter}}

\begin{document}

\title{Letters for Andrei:\ QBism and the Unfinished Nature of Nature}
\author{Christopher A. Fuchs\medskip
\\
\normalsize Department of Physics, University of Massachusetts Boston
\\
\normalsize 100 Morrissey Boulevard, Boston MA 02125, \ USA}

\date{2 September 2021}

\maketitle

\bigskip


\medskip

\begin{flushright}
\baselineskip=13pt
\parbox{4.0in}{
The import of the difference between pragmatism and rationalism is
now in sight throughout its whole extent. The essential contrast is
that {\it for rationalism reality is ready-made and complete from all
eternity, while for pragmatism it is still in the making, and awaits
part of  its complexion from the future}. On the one side the
universe is absolutely secure, on the other it is still pursuing its
adventures.\medskip}\\
--- William James\bigskip
\end{flushright}

\begin{flushright}
\baselineskip=13pt
\parbox{4.2in}{
Are we not obliged to admit that more or less ``measurement-like'' processes are going on more or less all the time more or less everywhere?\medskip}\\
--- John S. Bell, suitably reinterpreted for QBist purposes \bigskip
\end{flushright}

\section{Dedication}

Never has there been such a friend to the open and liberal discussion of quantum foundations as Andrei Khrennikov.  At his yearly meeting in {\Vaxjo}---the longest running series of conferences on the subject in history!---one will find all kinds, from the crackpot to the philosopher to the Nobel prizewinner.  Andrei's philosophy has always been that everyone is due respect in the forum: For everyone is united in wanting to crack the code of the quantum.  This has built the most wonderful atmosphere, and everyone knows it---it is why so many of us come back from year to year, from the early years when Andrei could fund our flights, to the {\it many, many more years after that\/} when he could not!  There is something about the whole experience of the meetings, not just the time in the lecture halls: from the long twilights and ubiquitous squawks of the seagulls, to the beautiful surrounding lakes of Sm{\aa}land, the thoughts that flow so spontaneously on the long walks between hotel and university, the deep and surprisingly coherent conversations till late at night in the Bishop's Arms, and the camaraderie of the banquets.

It is little known, and who would guess it, but those {\Vaxjo} meetings and Andrei himself played significant roles in the development of QBism~\cite{FuchsStacey2018,Fuchs2017,Fuchs2016}.  Andrei might even be called a reluctant godfather to QBism.  It all started when he asked me to be a co-organizer for his 2001 meeting.  I saw it as a watershed opportunity. Essentially {\it using\/} his generosity---``You are really American!,'' Andrei wrote me in an email---I gathered up quite a number of friends from the quantum-information world (Asher Peres, Danny Greenberger, David Mermin, Herb Bernstein, Carl Caves, R\"udiger Schack, Richard Jozsa, Ben Schumacher, John Smolin) to participate, as I felt that revolution was in the air: In the quantum-foundations world, it was just a matter of bringing some adults into the room---those who already understood that quantum states just are {\it information}.  I figured that would fix everything~\cite{Fuchs2001}. But how little I was prepared for my own transformation by the end of the meeting!

What was up was that in the interstices, Caves, Schack, and I used the opportunity to finish writing our first ``quantum Bayesian'' paper, ``Quantum Probabilities as Bayesian Probabilities''~\cite{CFS02}.  However as we did, I became and more and more suspicious of the consistency of our considerations.  In the end, I believe we presented little more than a facade of the Bayesianism we had hoped to inject into quantum theory. In fact, what had gotten in the way of our consistency was the very word ``information.''

The aftermath was a gut-wrenching experience as I started to understand that the usual conception of information had too much of a ``factive character''~\cite{Timpson08a} to be consistent with our hoped-for banishment of quantum states from reality.  Ultimately, the corrective was to abandon the road of E.~T. Jaynes's objective Bayesianism with its foundation resting on Cox's axioms~\cite{Jaynes2003}, and change course to Bruno de Finetti's radically subjective Bayesianism with its foundation of Dutch-book coherence~\cite{deFinetti1990}.  Only that concept was nimble enough to remove the factive character of probabilities and consequently the factive character of quantum states. Quantum states were not information; they were beliefs~\cite{Fuchs2002}!! That really was the start of the long road to QBism~\cite{Fuchs10a,Stacey2019a}.

Here is the way I put the process in the Introduction to my largest email samizdat\footnote{Samizdat is Russian for an underground publication, but I use the term here to denote one of my posted or published collections of emails.} (2,349 pages)  \cite{Samizdat2}:
\bq
\noindent The remediation of these problems required steps more radical than I ever imagined I could or would take.  For instance, I came to recognize that even probability-1 assignments must be personalist Bayesian probabilities of a cloth with any other Bayesian probability.  Moreover, I started to understand that not only quantum states, but the operators used for describing measuring devices and all quantum time evolution maps as well, even Hamiltonian evolutions, must be viewed in strictly personalist Bayesian terms.  If these steps were not taken, the view of quantum theory we had been striving to construct would simply topple from inconsistency.  But it was no easy pill for even my most sympathetic colleagues and collaborators to swallow.
\eq
Nearly everyone I encountered thought I had gone off the deep end: {\it That I had completely given up on the concept of reality}.  Certainly the participants at Andrei's meetings did.  (For instance, after Ted H\"ansch's talk on QBism during the 2015 meeting, Avshalom Elitzur stood up and declared angrily, ``You cannot tell me that I am a figment of Chris Fuchs's imagination!'')

Interestingly though I saw the whole process in the opposite light:  I saw the method of seeking absolute consistency between our interpretations of the quantum symbols and de Finetti's radically subjective notion of probability as a powerful method for homing in on an {\it ontology\/} to go with quantum theory---i.e., a positive statement about the character and contents of the world.  That's not giving up on reality, but a way of protecting oneself from being seduced by false prophets---you know, like the prophets of nonlocality or pilot waves or the wavefunction of the universe.

On the 9th of October 2001, just four short months after the {\Vaxjo} meeting that got all this started, I wrote to David Mermin:
\bq
\noindent Collecting it up, it's hard to believe I've written this much in the
little time since {\Vaxjo}. I guess it's been an active time for
me. I think there's no doubt that I've gone through a phase
transition. For all my Bayesian rhetoric in the last few years, I
simply had not realized the immense implications of holding fast to
the view that ``probabilities are subjective degrees of belief.'' Of
course, one way to look at this revelation is that it is a {\it
reductio ad absurdum\/} for the whole point of view---and that will
certainly be the first thing the critics pick up on. But---you
wouldn't have guessed less---I'm starting to view it as a godsend.
For with this simple train of logic, one can immediately stamp out
the potential reality/objectivity of any of a number of terms that
might have clouded our vision.  With so much dead weight removed, the little part left behind may finally have the strength to
support an ontology.
\eq
In fact I posted one year's worth (229 pages) of all the painful thinking that went into my new radical stance in a mini-samizdat titled, {\sl Quantum States:\ What the Hell Are They?\ The Post-{\Vaxjo} Phase Transition}~\cite{FuchsBreak}.

But who in their right mind would publish their emails!?  Well, I tell my reasons for it in the Introduction to my first samizdat, {\sl Notes on a Paulian Idea:\ Foundational, Historical, Anecdotal \& Forward-Looking Thoughts on the Quantum}~\cite{NPI1}, posted as {\tt arXiv:quant-ph/0105039} on the first anniversary of the Cerro Grande Fire that swept through Los Alamos, New Mexico when I was a postdoc there. But that brings us back to Andrei and his wonderful liberal ways.

The story goes that Andrei read the abstract of the posting, thought ``this sounds interesting,'' and hit the print button. He hadn't noticed that the pdf was 504 pages long.  When he later entered the printing room, he discovered that the printer couldn't handle the stack and there were disheveled pages all over the floor!  Only then did he realize what he had gotten himself into.  What happened next I'm not completely sure of, but it appears from my email record that while visiting Alexander Holevo and me at Bell Labs in October 2001, Andrei proposed to use his resources at {\Vaxjo} University (now Linnaeus University) to print a number of paperback copies of the samizdat.  There followed a printing of 100 copies, and then later 100 more~\cite{NPI2}.  Later still, I successfully approached Cambridge University Press with the idea of a hardbound edition of the book ({\sl Coming of Age with Quantum Information:\ Notes on a Paulian Idea}, 2010, \cite{Fuchs2010}), but I would have never reaped that reward or the eight excellent reviews the book received \cite{Greenberger2011,Cavalcanti2011,Trabesinger2011,Benedictus2012,Cavalcanti2012,Dick2013,Morikoshi2015,Bilban2020} without the confidence Andrei instilled in me through his original kindness.

And therein lies the paradox of Andrei, or what I called earlier his wonderful liberal ways.  Andrei has never yet made the step to thinking of QBism as the most appropriate interpretation for {\it quantum physics}, but that has never stopped his support for those of us developing the idea.  Witness the book mentioned above, but it goes deeper than that, as evidenced at every {\Vaxjo} meeting since 2001.  At Andrei's instigation (he would often call attention to QBism in his opening remarks), the debate would start right away.  So many tomatoes were thrown at QBism for sure, but learning how to dodge them sharpened our concepts. Certainly QBism was never ignored.  Compare that attitude with something the scholar Travis Norsen recently posted on social media, ``Are there any serious people that take QBism seriously? It is so obviously empty and stupid. \ldots\ [T]here is a real danger, I think, of providing undeserved oxygen to something that is, in fact, stupid and empty, by talking about it too much and too seriously.'' At most ``philosophy of physics'' conferences (so called), QBism is simply dead in the water.\footnote{Thank God however that ``standardized'' philosophy of physics (I'm searching for the right insult) is not the only game in town. Recently a number of philosophers of a phenomenological bent have started a constructive dialogue with QBism.  Witness for instance the forthcoming volume, {\sl Phenomenology and QBism:\ New Approaches to Quantum Mechanics}, edited by Philipp Berghofer and Harald A. Wiltsche (Routledge, New York, 2022). In the meantime see: \cite{Fuchs2021,Wiltsche2020,Bitbol2020a,Tremblaye2020,Bitbol2020b,Berghofer2020,Boge2021,Pienaar2022,Boge2022,French2022,Bitbol2022,Baeyer2022}.} But I think Andrei has taken a kind of pride in recognizing that there is something genuinely to be developed in QBism, even if it is adjacent to his own concerns.  On 14 August 2003, he wrote me, ``Hej Chris! I just like science as it is. \ldots\ You know that many people have been in {\Vaxjo}, but I really see that you have a huge potential for development.''

So, how shall I tribute Andrei in this volume?  With an email collection of course!\footnote{It doesn't help that besides the publication of my book, Andrei also allowed me to submit two email collections to his conference proceedings~\cite{Fuchs02b,Fuchs07a}---the precedent has already been set!} But with what theme?  This time it ought to be something big.  Here goes.

One of the troubles of QBism's ontological program is that it is so sideways to the ways most run-of-the-mill philosophers of physics (say, Norsen, Tim Maudlin, David Albert, Michael Esfeld, and others of similar ilk) think, they don't even have the tools to parse its sentences.  They simply can't see it as having to do with ontology at all.  Maybe there is no remedy for this except to wait for the generation to die away.  But QBism can and must move forward.  To this end, I have tried to select some emails\footnote{Mildly edited.} for the current collection that might inspire a young researcher to throw in on the QBist ontological project, to help develop it on its own terms. The road to a proper, detailed QBist ontology is sure to be a hard one, but one has to start somewhere. Thus, please take these pages as loose notes or supplements to the more formal QBist writings touching on the ontological project in the past~\cite{FuchsStacey2018,Fuchs2017,Fuchs2016,Fuchs2021,Pienaar2022,Fuchs2014,Pienaar2020,Pienaar2021}.  At the very least, I hope these exchanges will give some sense of the ways QBists think in contrast to the ways those philosophers do who are happy enough with a (cheaply found) ``fake,'' as Schr\"odinger put it~\cite{Schroedinger54}:
\bq
\noindent
In an honest search for knowledge you quite often have to abide by ignorance for an indefinite period.  Instead of filling a gap by guesswork, genuine science prefers to put up with it; and this, not so much from conscientious scruples about telling lies, as from the consideration that, however irksome the gap may be, its obliteration by a fake removes the urge to seek after a tenable answer.  So efficiently may attention be diverted that the answer is missed even when, by good luck, it comes close at hand.  The steadfastness in standing up to a {\it non liquet}, nay in appreciating it as a stimulus and a signpost to further quest, is a natural and indispensable disposition in the mind of a scientist.  This in itself is apt to set him at variance with the religious aim of closing the picture, unless each of the two antagonistic attitudes, both legitimate for their respective purposes, is applied with  prudence.
\eq

So, here is to you Andrei.  Thank you for tolerating QBism's many {\it non liquets\/} past and never giving up hope that one day one of us will crack the code of the quantum.

\section{5 Jan.\ 2009 \ \ {\it What I Want Out of Pauli-Fierz} \ \ (to H.~C. von Baeyer \& D.~M. Appleby)}

Thank you both for the great New Year's reading gift of the last couple of days.  It's funny how these things work out; it arrived at just the right time for me.  For, I was able to put it in juxtaposition with my reading of the latest issue of {\sl William James Studies}.  The two together caused me to think very much about how I want to frame the coming year.  And it struck me as worthwhile to try to record as clearly as I can at the moment what I am {\it personally\/} seeking to get out of a better knowledge of the byways explored by Pauli and Fierz (and potentially Jung).  There is no implication in this that this is what you {\it should\/} be seeking as well---it is only an effort to tie the strands of my life together a little better.

Here is what it is really all about for me---it is the root of the root and the bud of the bud, as e~e cummings put it.  A good vehicle for setting up what I want to say is one of Hans's passages:
\bq
\noindent Pauli summarily dismissed two extreme attitudes -- total separation of science from religion, and complete surrender to mystical experience.  The former approach was advocated in our times by the late paleontologist Stephen Jay Gould, who coined the phrase ``nonoverlapping magisteria'' for the respectful noninterference of the realms of nature and of morality, of what is and what should be.   Discussing a book by a German physicist who also believed in the separateness of science and religion, Pauli had once written to Fierz that he was appalled, and continuing:
\bq
\noindent ``[The book] is a reversion to the 19th century when religion and science lived in separate sections of the human soul -- politely exchanging greetings at a distance, while continually reassuring each other that they had nothing to do with each other -- and when the soul seemed to reside outside the boundaries of science.''
\eq
For Pauli it was obvious that science should be able to deal with the soul, and that the soul in turn can inform science.
\eq
in conjunction with one of Marcus's comments thereafter:
\bq
\noindent It is certainly true that I attach more weight to the opus than I do to quantum mechanics.  That is I am interested in quantum mechanics because I am interested in the opus, rather than the other way round.  For Chris it is, perhaps, a little different.  Though I don't think it is all that different, actually. It is true that Chris doesn't use the language of souls (at least I can't recall him doing so in my hearing).  But he is deeply concerned with, for example, the question of human freedom.  He will have to speak for himself, but I believe that for him too his interest in quantum mechanics is secondary to other considerations.
\eq
For it is true that I rarely speak of ``souls,'' ``religion,'' or ``redemption.''  These terms are mostly dead terms for me---they don't stir my soul, so to speak---or maybe I simply don't understand them well enough yet to see their ultimate usefulness for what I {\it do\/} want to get at.  (Much like I have never understood what the search for ``elegance'' can possibly mean when it comes to forming physical theories, say, as a criterion for string theory:  it is a term that is dead to me.)  It is maybe in this way, or more carefully, {\it in this detail}, that I part company from your tentative feelings on the opus.

{\it Nonetheless}, there is no doubt that I believe there is a place---a very important place---for humanistic concerns within physics proper.  It seems to me it goes to the core of what quantum mechanics is trying to tell us.  You'll find the point made over and over in my ``Activating Observer'' resource-material document, which both of you have versions of.  But I thought the point was made very nicely in the {\it setting of pragmatism more generally\/} in this article that I was reading at the time your emails arrived---``\,`The Many and the One' and the Problem of Two Minds Perceiving the Same Thing'' by Mark Moller in Vol.~3 of {\sl William James Studies}~\cite{Moller2008}:
\bq
Each of these claims about reality is crucial to James's attempt to offer an alternative to the metaphysical theories of the absolute idealists. The importance of the claim that reality is continuous and in flux is that it offers an alternative to the absolute idealists' view that the universe is ``known by one [infinite] knower in one act, with every feature preserved, and every relation apprehended.'' This means that for the absolute idealists, the universe is forever fixed so as to make real change impossible. We make no difference in such a universe. We neither improve upon it through our efforts nor make it worse. James rejects this view completely. His aim is to argue for a conception of the universe that allows real change to occur in it and where our efforts have a role in causing it. He goes on in the passage from {\sl The Many and the One\/} manuscripts quoted above to make his point:
\bq
\noindent
    This picture of the irremediably pluralistic evolution of things, achieving unity by experimental methods, and getting it in different shapes and degrees and in general only as a last result, is what has made me give to my volume the title of {\sl The Many and the One}.
\eq
According to James, we, as conscious agents in the universe, have an active role in introducing new content and unity into it. Such a view thus aligns his radical empiricism with his meliorism. In earlier essays, eventually published together as {\sl The Will to Believe\/} (1897), and in lectures that he gave to teachers, eventually published as {\sl Talks to Teachers on Psychology\/} (1899), James took the position against the absolute idealists that the ultimate fate of the universe has yet to be decided. He insisted that it is an open question as to whether evil will triumph over good or the other way around. This, in turn, led him to claim that our choices and actions do make a difference in the universe, and, in fact, a crucial one. They help to decide how ``the everlasting battle of the powers of the light with those of darkness'' will turn out. This melioristic attitude only makes sense if the universe is malleable to human action, and, thus, it is one of James's aims in developing his metaphysics to explain how this malleability is possible.
\eq
The very phrases ``our choices and actions do make a difference in the universe'' and ``this melioristic attitude only makes sense if the universe is malleable to human action'' mean outright that there must be room for a humanistic element of some sort within physics itself.

Here's the way I put it to Lucien and Joy Christian when I was in a poetic mood last year:
\bq
\noindent
My view of the universe is that it is many---that it ultimately cannot be unified, for it is alive and changing and creative in a very deep sense.  Moreover, that reality is, to some not-yet well-understood extent, plastic:  It can be molded by our actions.  Thus, though humanity is quite well a Darwinian accident, now that it is here, it is a significant component of the universe that must be reckoned with.
\eq
Finally, it seems worthwhile for me to let James say it himself (from his article ``Pragmatism and Humanism''). It captures with his beautiful sweep the romantic thought that really keeps me going from day to day:
\bq
In many familiar objects every one will recognize the human element.
We conceive a given reality in this way or in that, to suit our
purpose, and the reality passively submits to the conception. You can
take the number 27 as the cube of 3, or as the product of 3 and 9, or
as 26 {\it plus\/} 1, or 100 {\it minus\/} 73, or in countless other
ways, of which one will be just as true as another. You can take a
chess-board as black squares on a white ground, or as white squares
on a black ground, and neither conception is a false one.

You can treat the adjoined figure as a star, as two big triangles
crossing each other, as a hexagon with legs set up on its angles, as
six equal triangles hanging together by their tips, etc. All these
treatments are true treatments---the sensible {\it that\/} upon the
paper resists no one of them. You can say of a line that it runs
east, or you can say that it runs west, and the line {\it per se\/}
accepts both descriptions without rebelling at the inconsistency.

We carve out groups of stars in the heavens, and call them
constellations, and the stars patiently suffer us to do so,---{\it
though\/} if they knew what we were doing, some of them might feel
much surprised at the partners we had given them. We name the same
constellation diversely, as Charles's Wain, the Great Bear, or the
Dipper. None of the names will be false, and one will be as true as
another, for all are applicable.

In all these cases we humanly make an {\it addition\/} to some
sensible reality, and that reality tolerates the addition. All the
additions `agree' with the reality; they fit it, while they build it
out. No one of them is false. Which may be treated as the {\it
more\/} true, depends altogether on the human use of it. If the 27 is
a number of dollars which I find in a drawer where I had left 28, it
is 28 minus 1. If it is the number of inches in a board which I wish
to insert as a shelf into a cupboard 26 inches wide, it is 26 plus 1.
If I wish to ennoble the heavens by the constellations I see there,
`Charles's Wain' would be more true than `Dipper.' My friend
Frederick Myers was humorously indignant that that prodigious
star-group should remind us Americans of nothing but a culinary
utensil.

What shall we call a {\it thing\/} anyhow? It seems quite arbitrary,
for we carve out everything, just as we carve out constellations, to
suit our human purposes. For me, this whole `audience' is one thing,
which grows now restless, now attentive. I have no use at present for
its individual units, so I don't consider them. So of an `army,' of a
`nation.' But in your own eyes, ladies and gentlemen, to call you
`audience' is an accidental way of taking you. The permanently real
things for you are your individual persons. To an anatomist, again,
those persons are but organisms, and the real things are the organs.
Not the organs, so much as their constituent cells, say the
histologists; not the cells, but their molecules, say in turn the
chemists.

We break the flux of sensible reality into things, then, at our will.
We create the subjects of our true as well as of our false
propositions.

We create the predicates also. Many of the predicates of things
express only the relations of the things to us and to our feelings.
Such predicates of course are human additions. Caesar crossed the
Rubicon, and was a menace to Rome's freedom. He is also an American
schoolroom pest, made into one by the reaction of our schoolboys on
his writings. The added predicate is as true of him as the earlier
ones.

You see how naturally one comes to the humanistic principle: you
can't weed out the human contribution. Our nouns and adjectives are
all humanized heirlooms, and in the theories we build them into, the
inner order and arrangement is wholly dictated by human
considerations, intellectual consistency being one of them.
Mathematics and logic themselves are fermenting with human
rearrangements; physics, astronomy and biology follow massive cues of
preference. We plunge forward into the field of fresh experience with
the beliefs our ancestors and we have made already; these determine
what we notice; what we notice determines what we do; what we do
again determines what we experience; so from one thing to another,
altho the stubborn fact remains that there is a sensible flux, what
is {\it true of it\/} seems from first to last to be largely a matter
of our own creation.

We build the flux out inevitably. The great question is: does it,
with our additions, {\it rise or fall in value}? Are the additions
{\it worthy\/} or {\it unworthy}? Suppose a universe composed of
seven stars, and nothing else but three human witnesses and their
critic. One witness names the stars `Great Bear'; one calls them
`Charles's Wain'; one calls them the `Dipper.' Which human addition
has made the best universe of the given stellar material? If
Frederick Myers were the critic, he would have no hesitation in
`turning down' the American witness.

Lotze has in several places made a deep suggestion. We na\"{\i}vely
assume, he says, a relation between reality and our minds which may
be just the opposite of the true one. Reality, we naturally think,
stands ready-made and complete, and our intellects supervene with the
one simple duty of describing it as it is already. But may not our
descriptions, Lotze asks, be themselves important additions to
reality? And may not previous reality itself be there, far less for
the purpose of reappearing unaltered in our knowledge, than for the
very purpose of stimulating our minds to such additions as shall
enhance the universe's total value. {\it `Die Erh\"ohung des vorgefundenen Daseins'\/} is a phrase used by Professor Eucken
somewhere, which reminds one of this suggestion by the great Lotze.

It is identically our pragmatistic conception. In our cognitive as
well as in our active life we are creative. We {\it add}, both to the
subject and to the predicate part of reality. The world stands really
malleable, waiting to receive its final touches at our hands. Like
the kingdom of heaven, it suffers human violence willingly. Man {\it
engenders\/} truths upon it.

No one can deny that such a role would add both to our dignity and to
our responsibility as thinkers. To some of us it proves a most
inspiring notion. Signore Papini, the leader of Italian pragmatism,
grows fairly dithyrambic over the view that it opens of man's
divinely-creative functions.

The import of the difference between pragmatism and rationalism is
now in sight throughout its whole extent. The essential contrast is
that {\it for rationalism reality is ready-made and complete from all
eternity, while for pragmatism it is still in the making, and awaits
part of  its complexion from the future}. On the one side the
universe is absolutely secure, on the other it is still pursuing its
adventures.

We have got into rather deep water with this humanistic view, and it
is no wonder that misunderstanding gathers round it. It is accused of
being a doctrine of caprice. Mr.\ Bradley, for example, says that a
humanist, if he understood his own doctrine, would have to `hold any
end, however perverted, to be rational, if I insist on it personally,
and any idea, however mad, to be the truth if only some one is
resolved that he will have it so.' The humanist view of `reality,' as
something resisting, yet malleable, which controls our thinking as an
energy that must be taken `account' of incessantly (tho not
necessarily merely {\it copied}) is evidently a difficult one to
introduce to novices. The situation reminds me of one that I have
personally gone through. I once wrote an essay on our right to
believe, which I unluckily called the {\it Will\/} to Believe. All
the critics, neglecting the essay, pounced upon the title.
Psychologically it was impossible, morally it was iniquitous. The
`will to deceive,' the `will to make-believe,' were wittily proposed
as substitutes for it.

{\it The alternative between pragmatism and rationalism, in the shape
in which we now have it before us, is no longer a question in the
theory of knowledge, it concerns the structure of the universe
itself.}

On the pragmatist side we have only one edition of the universe,
unfinished, growing in all sorts of places, especially in the places
where thinking beings are at work.

On the rationalist side we have a universe in many editions, one real
one, the infinite folio, or {\it \'edition de luxe}, eternally
complete; and then the various finite editions, full of false
readings, distorted and mutilated each in its own way.
\eq

My anti-Platonist tendencies certainly come out in my feeling of agreement with this quote.  There is no sympathy in it to the idea that one might be {\it attuned\/} with a divine mind privy to ``THE way'' the universe is constructed---which as Marcus, I think, rightly points out, seems to have been Einstein's modus operandi when theorizing.  At most, one might be attuned to the content of this other quote of James:
\bq
If you follow the pragmatic method, you cannot look on any [theory] as closing your quest.  You must bring out of each [theory] its practical cash-value, set it at work within the stream of your
experience.  It appears less as a solution, then, than as a program
for more work, and more particularly as an indication of the ways in
which existing realities may be {\it changed}.

{\it Theories thus become instruments, not answers to enigmas, in
which we can rest.}  We don't lie back upon them, we move forward,
and, on occasion, make nature over again by their aid.
\eq
Physical theories, by this view, are conceptual means and tools for making change in the world.\footnote{This is not the dry instrumentalism one incessantly hears of in philosophy of science circles---that an instrumentalist is one who believes a scientific theory is {\it solely\/} an instrument for {\it prediction}.  That is the term of insult they usually throw at Bohr, our quantum Bayesianism, and similar efforts: ``it is merely instrumentalism,'' they say.  (See Asher Peres's funny story at the beginning of~\cite{Peres2005}.)  But prediction only takes a secondary role here, at best.  What's being talked about instead is a far richer concept than those guys---Jeff Bub, Harvey Brown, David Albert, and the like---are even aware of.  For this kind of ``instrumentalism'' (if it should even be called that) the image to associate with a theory is not so much Babbage's analytical engine, but a hammer and a socket set.}  Can one be attuned to such a thing in the Platonic sense that Marcus explores?  Perhaps.  But if so, it's not the medieval Platonism Marcus was talking about---it is instead being attuned to how best be an agent of change.  (Perhaps the exemplar of this modified Platonism would be Barack Obama instead of Einstein \ldots\ so that a pragmatically modified Platonist would look on him and say, ``Now, that man is someone who feels the Old One!'')

In any case, what's important here is that I genuinely do see a humanistic role in the very maintenance of the universe.  And all of this is potentially independent, and certainly broader, than considerations to do with quantum mechanics.  The universe is partially powered by the inhuman, to be sure.  But it is also partially powered by \underline{\it belief}---that I do believe full well.  And to the extent that religion or religious feeling are sources of belief, they do indeed play a role in the very construction of the universe.  However, from this point of view religion is a special case of something much bigger---namely, belief generally.

I'll quote James one last time, so that you get a precise sense of what I mean here, but then I'll get back to the quantum and Pauli/Fierz and tell you way I'm saying all this.  From ``The Sentiment of Rationality'':
\bq
Now, I wish to show what to my knowledge has never been clearly
pointed out, that belief (as measured by action) not only does and
must continually outstrip scientific evidence, but that there is a
certain class of truths of whose reality belief is a factor as well
as a confessor; and that as regards this class of truths faith is not
only licit and pertinent, but essential and indispensable. The truths
cannot become true till our faith has made them so.

Suppose, for example, that I am climbing in the Alps, and have had
the ill-luck to work myself into a position from which the only
escape is by a terrible leap. Being without similar experience, I
have no evidence of my ability to perform it successfully; but hope
and confidence in myself make me sure I shall not miss my aim, and
nerve my feet to execute what without those subjective emotions would
perhaps have been impossible. But suppose that, on the contrary, the
emotions of fear and mistrust preponderate; or suppose that, having
just read [W.~K. Clifford's] {\sl Ethics of Belief}, I feel it would be sinful to act upon an assumption unverified by previous experience---why, then
I shall hesitate so long that at last, exhausted and trembling, and
launching myself in a moment of despair, I miss my foothold and roll
into the abyss. In this case (and it is one of an immense class) the
part of wisdom clearly is to believe what one desires; for the belief
is one of the indispensable preliminary conditions of the realization
of its object. {\it There are then cases where faith creates its own
verification}. Believe, and you shall be right, for you shall save
yourself; doubt, and you shall again be right, for you shall perish.
The only difference is that to believe is greatly to your advantage.

The future movements of the stars or the facts of past history are
determined now once for all, whether I like them or not. They are
given irrespective of my wishes, and in all that concerns truths like
these subjective preference should have no part; it can only obscure
the judgment. But in every fact into which there enters an element of
personal contribution on my part, as soon as this personal
contribution demands a certain degree of subjective energy which, in
its turn, calls for a certain amount of faith in the result---so
that, after all, the future fact is conditioned by my present faith
in it---how trebly asinine would it be for me to deny myself the use
of the subjective method, the method of belief based on desire!

In every proposition whose bearing is universal (and such are all the
propositions of philosophy), the acts of the subject and their
consequences throughout eternity should be included in the formula.
If $M$ represent the entire world minus the reaction of the thinker
upon it, and if $M + x$ represent the absolutely total matter of
philosophic propositions ($x$ standing for the thinker's reaction and
its results)---what would be a universal truth if the term $x$ were
of one complexion, might become egregious error if $x$ altered its
character. Let it not be said that $x$ is too infinitesimal a
component to change the character of the immense whole in which it
lies imbedded. Everything depends on the point of view of the
philosophic proposition in question. If we have to define the
universe from the point of view of sensibility, the critical material
for our judgment lies in the animal kingdom, insignificant as that
is, quantitatively considered. The moral definition of the world may
depend on phenomena more restricted still in range. In short, many a
long phrase may have its sense reversed by the addition of three
letters, {\it n-o-t}; many a monstrous mass have its unstable
equilibrium discharged one way or the other by a feather weight that
falls.
\eq

These, however, are very big things and very big thoughts James is speaking of.  And even though they stir my heart, they remain too vague to transform science as a whole (and indeed the world) in the way that I hope we'll one day transform it.  I love the sound of what I'm hearing, but at the end of the day, I'm still not completely sure what I {\it am\/} hearing.  How do I know I'm not simply fooling myself with pretty words?  The main point I want to make at this juncture is that these ideas which drive me forward---these still-too-vague ideas---are {\it supra}- quantum mechanics.  It is 19$^{\rm th}$ century philosophy, not physics.

What now is the role of quantum mechanics within this system of thought I'm laying out?  It is that it is a miniature version of these general points.  {\it But}, though it is a miniature version, it is an extremely {\it precise\/} version!  And that's its ace in the hole, as Hans said of Pauli's discussion with Huxley.  Here's where I agree with Marcus,
\bq
\noindent Quantum mechanics is important because it is as close to a refutation of the classical world picture as one could hope to get.  It is very rare that something gets refuted in science as completely and finally as the classical hypothesis, of the world machine, has been refuted.
\eq
Thus, quantum mechanics is a precision laboratory for defining and testing these things {\it we think we see\/} on the horizon.  When we've got it in quantum mechanics, we know we have it.

This finally is where I can express the value for me of learning more about the Pauli--Fierz and the Pauli--Jung correspondence.  There is an amazing amount of development of the metaphysics of a malleable world in the writings of William James and F.~C.~S. Schiller\footnote{F.~C.~S. should not be confused with the famous Schiller, namely Friedrich.  F.~C.~S. stands for Ferdinand Canning Scott---a mostly forgotten pragmatist, who was one of the deepest of the lot if you ask me.}.\footnote{And no disrespect intended, but Pauli and Fierz have got nothing on the sheer volume of even James's and Schiller's correspondence on these subjects, much less their published works.  Pauli was an amateur in comparison \ldots\ as was necessarily the case, being a professional physicist most hours of the day.}  And lo and behold, in which direction does that development lead?  In significant parts at least, it leads in quite the same direction as Pauli's metaphysic.  I don't know that I've ever emphasized this to you two.  When James, for instance, asks the question, ``What are the materials of the universe's composition?,'' the answer he tries to develop is that they are things {\it neutral\/} to the material/mental or physical/psychical distinction.  In fact, he saw the material and mental as {\it complementary}, but exclusive, aspects of the basic stuff.  So, you see the similarity.\footnote{At this point, I'd definitely encourage you to read the whole of the Moller paper mentioned above.  It gives quite a decent summary of the Jamesian metaphysic.  Moreover, in it not only will you see a resemblance between James's ideas and Pauli's ``neutral language'' considerations, but also, in the objections to James's pluralism, you will find some difficulties that we quantum Bayesians (who take an ``alchemical view'' on quantum measurement) must eventually address.}

But James and Schiller knew no quantum mechanics of course.  Particularly, they were not privy to this precision development laboratory for their metaphysic that Pauli and Fierz knew so well.  Therefore, I need, I crave, to know more of the precise things Pauli and Fierz were talking about for just this reason.  The suspicion is that it will inspire us and help us connect the dots between the quantum formalism and the bigger, much richer, and far-from-really-developed idea of a malleable/alchemical world.

Marcus says,
\bq
\noindent Quantum mechanics, as it stands now, is little more than a set of calculational procedures.  The calculations show that the vision of the world-machine is completely without empirical foundation.  However, the mechanical vision of things has not been replaced with any other.
\eq
of which I have a minor qualm, but I think he is right in further saying, ``The classical vision of things \ldots\ was deeply wrong about the relation of mind to matter.'' For I think this neutral stuff that Pauli and James speak of---James called it ``pure experience'' but I am not completely happy with that term---will indeed be the ``shocking'' ingredient that will replace the mechanical vision of things.  So, it is not that we have not started the process of replacement; it is only that it has been slow going.

Hans, I suspect, has wondered why I place so much emphasis on wanting to know the most, within all of Pauli's non-$x$-ing writings and correspondence, about the part to do with the idea of detached and nondetached observers (activating observers).  It is because, as I see it, a thorough study of what it means for an observer or agent to be nondetached from the phenomena he helps bring about is the very starting point for a development of James and Pauli's neutral-stuff ontology.  If we cannot get the activating-observer idea right and understand all its facets, then I feel we have no chance whatsoever of developing a full-blown ``neutral monism''\footnote{Neutral monism is what it is sometimes called in the literature, but a much better name for the idea in the Jamesian context would be ``neutral pluralism.''  Until ten seconds after writing this, I thought I was the first person to make up this term.  But apparently Ruth Anna Putnam beat me to it (damned the power of this internet; it is the second time it has foiled me today):
\bq
[Putnam writes:] Another key element of James's radical empiricism is his rejection of mind/matter dualism as well as its reduction to either materialism or idealism.  In its place, he offers---it is the title of one of his essays---a world of pure experience.  In that world consciousness {\it as an entity\/} does not exist.  But neither is consciousness a function of matter, for matter {\it as an entity\/} also does not exist.  Ultimately there are only pure experiences (and, perhaps, experienceables---that is a difficult interpretative question), experiences which only in retrospect are {\it taken\/} either as part of a stream of thought or as physical objects.  Although one is tempted to call this view a neutral monism, it is, in my opinion, more properly thought of as a neutral pluralism---neutral in not favoring either thought or matter, plural because ``there is no {\it general\/} stuff of which experience at large is made.  There are as many stuffs as there are `natures' in the things experienced \ldots\ and save for time and space (and, if you like, for `being') there appears no universal element of which all things are made.''
\eq
}.  It is the very place to start, for it forces the issue of this new ontology in its very statement.  What is an observer if not psychical?  What is an activator if not physical?  How can we really combine these two aspects of the phenomenon via the help of a Paulian neutral language?

Unfortunately in the technical development of this first step, Marcus and I have hit a great snag by not yet understanding the structure (a.k.a.\ knowing how to prove the existence of) these damned SICs.\footnote{SICs refers to the notion of the symmetric informationally complete quantum measurements.  For some updated references about this aspect of the QBist research project, see: \cite{DeBrota2021,DeBrota2020,Appleby2017,Appleby2015}. \label{ThatFootnote}}
The reason I say this is because I see no way forward to a more precise definition of the activating-observer concept than by a kind of ``generalized Newton's third law'' of the style I wrote you about last month.  And I see no way to make progress in that intermediate program without the ability to completely rewrite quantum mechanics (and particularly the Born rule) in terms of SICs.  For completeness, let me reproduce that little essay:
\bq
Everybody has their favorite speculation about what powers quantum
information and computing.  Some say it is the superposition
principle, some say it is the parallel computation of many worlds,
some say it is the mysteries of quantum entanglement, some say it is
the exponential growth of computational space due to the tensor
product.  For my own part though, my favorite speculation is that it
is Newton's Third Law:  For every action, there is an equal and
opposite reaction.  Indeed I sometimes wonder if the very essence of
quantum mechanics isn't just this principle, only carried through far
more consistently than Newton could have envisioned.  That is to say,
absolutely {\it nothing\/} is exempt from it.

What do I mean by this?  What might have been exempt from the
principle in the first place?  To give an answer, let me note an
equivalent formulation of old Newton.  For every {\it reaction}, there is an equal and opposite {\it action}.  Strange sounding, but there's nothing wrong with it, and more importantly, this formulation allows for the possibility of an immediate connection to information theory.  In particular, we should not forget how information gathering is
represented in the Shannon theory.  An agent has gathered
information---by the very definition of the process---when something
in his environment has caused him to {\it react\/} by way of revising a prior expectation $p(h)$ (for some phenomenon) to a posterior expectation $p(h|d)$ (for the same phenomenon).

When information is gathered, it is because we are reacting to the
stimulation of something external to us.  The great lesson of quantum
mechanics may just be that information gathering is physical.  Even
something so seemingly unimportant to the rest of the universe as the
reactions that cause the revisions of our expectations are not exempt
from Newton's Third Law.  When we react to the world's stimulations
upon us, it too must react to our stimulations upon it.

The question is, how might we envision a world with this
property---i.e., with such a serious accounting of Newton's law---but
in a way that does not make a priori use of the information gathering
agent himself?  If the question can be answered at all, the task of
finding an answer will be some tall order.  For never before in
science have we encountered a situation where the theorizing
scientist is so inextricably bound up with what he is trying to
theorize about in the first place.

It's almost a paradoxical situation.  On the one hand we'd like to
step outside the world and get a clear view of what it looks like
without the scientist necessarily in the picture.  But on the other
hand, to even pose the question we have to imagine an information
gathering agent set in the middle of it all.  You see, neither
Shannon nor any of modern information theory has given us a way to
talk about the concept of information gain without first introducing
the agent-centered concept of an expectation $p(h)$.

So, how to make progress?  What we do know is that we actually are in
the middle of the world thinking about it.  Maybe our strategy ought
to be to use that very vantage point to get as close as we can to the
goal.  That is, though we may not know what the world looks like
without the information gathering agent in it, we certainly do know
something about what it looks like with him in:  We know, for
instance, that he ought to use the formal structure of quantum
mechanics when thinking about physical systems.  Beyond that, we know
of an imaginary world where Newton's Third Law was never taken so
seriously:  It is the standard world of classical physics and
Bayesian probability.

Thus, maybe the thing to do first is to look inward, before looking
outward.  About ourselves, at the very least, we can ask how has the
formal structure of our {\it behavior\/} changed since moving from
what we thought to be a classical Bayesian world to what we now
believe to be a quantum world?  In that {\it differential}---the
speculation is---we may just find the cleanest statement yet of what
the quantum world is all about.  For it is in that differential, that
the world without us surely rears its head.

To do this, we must first express quantum mechanics in a way that it
can be directly compared to classical Bayesian theory, where the
information-gathering agent was detached from the world.
\eq

So, you see, there's work to be done all the way down to the SICs, and all the way up to a full-blown ontology of neutral pluralism and a malleable/alchemical reality.  In the case of the SICs, I'm actually more confident:  That problem is going to be solved if I have to recruit all of Hannibal's army.  But in the case of the ontology it's just verbal hints we have presently---there was only so much James, Schiller, Dewey, and others could discern from general premises, without detailed empirical input.\footnote{And only so much patience I have for straining through a set of thoughts framed so far away from direct quantum mechanical considerations.}  Clearly we will have to see our way through in that big task.  But Pauli and Fierz are the only two quantum physicists I know of who have already waded these waters to any depth.  If I could get hold of what little they may have already strained from the rocks below, it would be invaluable for this effort.  And that is what the title of this New Year's note is all about:  What I Really Want Out of a Pauli/Fierz-Correspondence Study.

Thank you both for not losing faith that there is something really important here.

\section{16 Oct.\ 2009 \ \ {\it The More and the Modest} \ \ (to L. Hardy)}

Greetings from [New Orleans].  This note is to say thanks again for your remarks at the panel discussion Friday~\cite{PD2009}.  You did something that none of the other ones did---you caused me to think and clarify, rather than piss me off from a display of overt closed-mindedness.  It amazes me the way Norsen and Valentini preach to me without understanding an ounce of the way I'm trying to see things.  (I just got another note from Norsen yesterday accusing me yet again of solipsism.)

Your point on Hilbert-space dimension on the other hand was a productive one, and thinking about how I should reply to you has helped me realize that my language needs a significant overhaul.  With better language, we inevitably get better understanding.  Here goes.

I used to ask, ``What is real of a quantum system?,'' and give ``Zing'' as a placeholder for the answer.  What is Zing?  The only thing I've been able to comfortably identify for it in the quantum formalism is Hilbert-space dimension.  But that certainly can't be enough to get the world going---or at least, I think you would (and did) say something like this.  John Sipe once wrote, ``the quantum world of this interpretation is a fixed, static thing.  It is a frozen, changeless place.''

That is, {\it at face value} there's not enough engine, not enough gears and pinions, in this bare ``reality'' to account for motion and change and indeterminism and novelty and growth and evolution and \ldots\ make the list just as big as you want \ldots\ all the things we see around us.   What is weird is that I agree.  Such a minimalist world doesn't even track with what I myself believe.  So it becomes a question of why I have doggedly resisted saying more.

I believe I have tracked down the troublemaker in my mode of expression, and it is this.  My trouble is {\it not\/} that I believe that there is {\it nothing\/} in the world, or perhaps a minimalist number of things (too minimal), as the solipsism-chargers seem so obsessed with accusing me of.  It is rather that my point of view admits {\it too many\/} things into the world---too many things of an independent and self-sustaining reality, things for which there are no equations; realities of which I am only willing to {\it point\/} to and say effectively, ``Yeah the world includes that too.''

This is because the ontology I imagine is the one of the humanist, rather than the trained or indoctrinated scientist.  Take Democritus as an example:  for him, the universe literally was ``atoms and void.''  That simple phrase was meant to account for all that is.  Similarly, the universe Einstein sought was literally meant to be a differentiable manifold supporting the solution to a very clever (though never found) differential equation---all else, even such as the flow of time, was simply illusion.  (Recall Mermin's talk at PIAF~\cite{Mermin2009}.)  The unifying theme in these two visions is that chairs and thumbs and bricks and all the like from our common experiences are secondary things---effective (operational) descriptions or illusions---having no real, primary existence of their own.  For a most relevant example, consider David Wallace's universe.  As far as I can tell, it is literally a single quantum state on a Hilbert space.  That vector, timelessly unchanging in a rotating frame, is the universe's whole substance.  The apparent indeterminism of the quantum world is subordinate to the greater monistic determinism and timelessness of the whole.

But here's the funny thing:  It is exactly of Einstein's and Wallace's universes, not my own, that I would say, ``the world of this interpretation is a fixed, static thing.  It is a frozen, changeless place.''  To my mind, these are both completely barren visions for the world.  I think our distaste for each other's proclivities comes from this:  It is a very old philosophical divide, that between the rationalist attitude and the empiricist attitude.  It has only become clear to me recently, but at my core I am an extreme empiricist.

Here is the way my friend William James put the distinction when he was in a relatively mild mood:
\bq
\noindent By empiricism I mean the tendency which lays most stress on the part, the element, the individual, treats the whole as a collection, and calls the universal an abstraction. By rationalism I mean the tendency to emphasize the universal, and to make the whole prior to the part, in the order both of logic and of being. The temper of rationalism is dogmatic: it willingly claims necessity for its conclusions.  Empiricism is more modest, and professes to deal in hypotheses only.
\eq
A more full-blooded statement of the issues involved, and just beautiful reading, can be found in a long quote I've plucked out for you from his essay ``The Sentiment of Rationality.''  See subsection at the end of this note.  If you find something worth reading in this note of my own composition, I hope you will take the time to read the longer quote of James.  In the end, it has become very crucial to my thinking.

Let me give a briefer hint of the issue here.  It has to do with the way James put the dangers of rationalism when he was in a fiercer mood:
\bq
Let me give the name of `vicious abstractionism' to a way of using concepts which may be thus described: We conceive a concrete situation by singling out some salient or important feature in it, and classing it under that; then, instead of adding to its previous characters all the positive consequences which the new way of conceiving it may bring, we proceed to use our concept privatively; reducing the originally rich phenomenon to the naked suggestions of that name abstractly taken, treating it as a case of `nothing but' that, concept, and acting as if all the other characters from out of which the concept is abstracted were expunged. Abstraction, functioning in this way, becomes a means of arrest far more than a means of advance in thought. It mutilates things; it creates difficulties and finds impossibilities; and more than half the trouble that metaphysicians and logicians give themselves over the paradoxes and dialectic puzzles of the universe may, I am convinced, be traced to this relatively simple source. {\it The viciously privative employment of abstract characters and class names\/} is, I am persuaded, one of the great original sins of the rationalistic
mind.
\eq

The way this is relevant to me is the following.  Far from thinking the world is an empty place, a place only with me in it.  I think it is full of things, overflowing with things.  {\it All\/} distinct things, from head to toe.  And literally so.  It is not a world made of six flavors of quarks glued together in various combinations.  It is not a world that maps to a single algorithm running on Rob Spekkens's favorite version of Daniel Dennett's mechanistic cellular automaton.  It is a world of heads and toes and doorknobs and dreams and ambitions and every kind of particular.  (And that is not a typo:  It is a world in which even dreams and ambition have substance.)  It is a world in which Vivienne Hardy is a distinct entity, not ``constructed'' of anything else, but a true-blue crucial piece of the universe as it is today---no less crucial than spacetime itself.

In modern parlance, I am not a reductionist.  And when my world is judged as empty (of all but me), I claim it is because I am being interpreted from the standpoint of an (explicit or implicit) reductionist worldview.  It is true that my envisioned world may be as cockamamie as James's heads of Borneo:
\bq
Taken as it does appear, our universe is to a large extent chaotic.  No one single type of connection runs through all the experiences that compose it.  If we take space-relations, they fail to connect minds into any regular system.  Causes and purposes obtain only among special series of facts.  The self-relation seems extremely limited and does not link two different selves together.  Prima facie, if you should liken the universe of absolute idealism to an aquarium, a crystal globe in which goldfish are swimming, you would have to compare the empiricist universe to something more like one of those dried human heads with which the Dyaks of Borneo deck their lodges.  The skull forms a solid nucleus; but innumerable feathers, leaves, strings, beads, and loose appendices of every description float and dangle from it, and, save that they terminate in it, seem to have nothing to do with one another.  Even so my experiences and yours float and dangle, terminating, it is true, in a nucleus of common perception, but for the most part out of sight and irrelevant and unimaginable to one another.
\eq
But it is not an {\it empty\/} world.

With this as a background, let me now return to Zing, Hilbert-space dimension, and the needed overhaul of my language.  I ask, ``What is real of a quantum system?'' and answer, ``its Hilbert-space dimension,'' but that is a very funny thing to say.  There is {\it so much\/} that is real of a quantum system, why would I ever say that?  I wouldn't say it of Vivienne Hardy, for instance,---I already declared this quite vocally---and she is a perfectly good example of a quantum system.  So, what am I really up to?

It is a bit like this:  Forget quantum mechanics, and think back to the days when the only physics known was basically Newtonian gravity.  Would any Newtonian have ever answered the question, ``What is real of a physical system?,'' with the declaration, ``Only its gravitational mass is real of it.''?  Most definitely not.  But that is because Newton merely/boldly taught us but a singular thing:  That every body in the universe, that every thing that can be carved out from it, had a previously undisclosed capacity---a numerical capacity to (try to) attract every other body in the universe.  On the one hand, I say ``merely'' because looked at in this way, it really is a very {\it modest\/} move---modest at least by the lights of the Steven-Weinberg--Stephen-Hawking-theory-of-everything generation:  It was no theory of everything---never pretended to be.  On the other hand, I say ``boldly'' because I certainly agree with my friend Hans Christian von Baeyer who wrote in one of his books~\cite{vonBaeyer2009}:
\bq
Great revolutionaries don't stop at half measures if they can go all the way.  For Newton this meant an almost unimaginable widening of the scope of his new-found law.  Not only Earth, Sun, and planets attract objects in their vicinity, he conjectured, but all objects, no matter how large or small, attract all other objects, no matter how far distant.  It was a proposition of almost reckless boldness, and it changed the way we perceive the world.
\eq

And therein lies the key, I think, to how I should modify my language.  When I ask myself what have we learned with quantum mechanics, I want to say first and foremost that we have learned how we should more consistently gamble upon the consequences of our interactions with external physical systems.  But why this new calculus for gambling?  Because quantum mechanics is uncovering that every object in the universe has a previously undisclosed {\it capacity}.

I would have liked to have said ``uncovered,'' but at this stage of research I must still settle for ``is uncovering''---it is an unfinished project to understand the significance of quantum mechanics in these terms.  Hilbert-space dimension, like gravitational mass, is representative of some universal capacity.  That's the real idea.  Dimension is a quality a body possesses in a way that, in a QBayesian understanding, it does not ``possess'' a quantum state.  If we conceptually delete an agent gambling upon the consequences of his interactions with a quantum system---the QBist says---we also conceptually delete its quantum state.  But there is no reason to believe the system and the system's capacity are deleted as well.

Capacity for what?  That's where the hard part begins.  Sometimes I say the capacity for birth and creation.  Sometimes I say the capacity to entertain counterfactuals.  Sometimes I say it is a capacity that can be harnessed to aid of computation, as in quantum computation.  The truth is, I don't yet know what I mean in any precise way.  I only know that I have a strong inner tug to thinking that the SIC calculus will help reveal a precise idea.

Anyway, I write all this to put the key idea into perspective and to practice a way of speaking that I had not used to any great extent before.  [Chris Timpson had pointed out that there are several similarities between the way I speak and the way the philosopher Nancy Cartwright speaks of ``capacities.''  And on this trip to New Orleans, I have confirmed it is very much so by starting to read her book, {\sl The Dappled World}.  So, at least I'm not alone in the world.]  You were indeed right to suspect that Hilbert-space dimension doesn't have enough gears and pinions to get things going, but neither did gravitational mass.  Nonetheless, the disclosing of that old capacity---gravitational mass---``changed the way we perceive the world'' (von Baeyer).  And so I think of our new capacity.

Thus I end this instalment by saying thanks again for pointing out a great deficiency in my choice of words and for consequently causing me to think.  I'll leave you with a sentence that I wrote in this file, but could not quite find a way to use in the essay.  It seems a shame to throw it away:
\begin{center}
Quantum mechanics calls out, ``I will not be a representative of your monistic dreams!''
\end{center}
When Norsen, Valentini, and Wiseman call me solipsist (or worse), I think it is because they are working from the middle of a monistic dream.

\subsection{Appendix: Passage from James's ``The Sentiment of Rationality''}

\subsubsection{The Part of Immediate Interest}

\bq
The facts of the world in their sensible diversity
are always before us, but our theoretic need is that
they should be conceived in a way that reduces their
manifoldness to simplicity. Our pleasure at finding
that a chaos of facts is the expression of a single
underlying fact is like the relief of the musician at
resolving a confused mass of sound into melodic or
harmonic order. The simplified result is handled
with far less mental effort than the original data; and
a philosophic conception of nature is thus in no
metaphorical sense a labor-saving contrivance. The
passion for parsimony, for economy of means in
thought, is the philosophic passion {\it par excellence};
and any character or aspect of the world's phenomena which gathers up their diversity into monotony
will gratify that passion, and in the philosopher's
mind stand for that essence of things compared with
which all their other determinations may by him be
overlooked.

More universality or extensiveness is, then, one
mark which the philosopher's conceptions must possess. Unless they apply to an enormous number of
cases they will not bring him relief. The knowledge
of things by their causes, which is often given as a
definition of rational knowledge, is useless to him
unless the causes converge to a minimum number,
while still producing the maximum number of effects.
The more multiple then are the instances, the more
flowingly does his mind rove from fact to fact. The
phenomenal transitions are no real transitions; each
item is the same old friend with a slightly altered
dress.

Who does not feel the charm of thinking that the
moon and the apple are, as far as their relation to the
earth goes, identical; of knowing respiration and
combustion to be one; of understanding that the
balloon rises by the same law whereby the stone
sinks; of feeling that the warmth in one's palm when
one rubs one's sleeve is identical with the motion
which the friction checks; of recognizing the difference between beast and fish to be only a higher
degree of that between human father and son; of
believing our strength when we climb the mountain
or fell the tree to be no other than the strength of
the sun's rays which made the corn grow out of
which we got our morning meal?

But alongside of this passion for simplification
there exists a sister passion, which in some minds---though they perhaps form the minority---is its rival.
This is the passion for distinguishing; it is the impulse to be {\it acquainted\/} with the parts rather than to
comprehend the whole. Loyalty to clearness and
integrity of perception, dislike of blurred outlines, of
vague identifications, are its characteristics. It loves
to recognize particulars in their full completeness,
and the more of these it can carry the happier it is.
It prefers any amount of incoherence, abruptness, and
fragmentariness (so long as the literal details of the
separate facts are saved) to an abstract way of conceiving things that, while it simplifies them, dissolves
away at the same time their concrete fulness. Clearness and simplicity thus set up rival claims, and make
a real dilemma for the thinker.

A man's philosophic attitude is determined by the
balance in him of these two cravings. No system
of philosophy can hope to be universally accepted
among men which grossly violates either need, or
entirely subordinates the one to the other. The fate
of Spinoza, with his barren union of all things in one
substance, on the one hand; that of Hume, with
his equally barren ``looseness and separateness'' of
everything, on the other---neither philosopher owning any strict and systematic disciples today, each
being to posterity a warning as well as a stimulus---show us that the only possible philosophy must be
a compromise between an abstract monotony and a
concrete heterogeneity. But the only way to mediate
between diversity and unity is to class the diverse
items as cases of a common essence which you discover in them. Classification of things into extensive ``kinds'' is thus the first step; and classification of their relations and conduct into extensive ``laws''
is the last step, in their philosophic unification. A completed theoretic philosophy can thus never be anything more than
a completed classification of the world's ingredients; and its
results must always be abstract, since the basis of every
classification is the abstract essence embedded in the living
fact---the rest of the living fact being for the time ignored by the
classifier. This means that none of our explanations are complete.
They subsume things under heads wider or more familiar; but the last
heads, whether of things or of their connections, are mere abstract
genera, data which we just find in things and write down.

When, for example, we think that we have rationally explained the
connection of the facts $A$ and $B$ by classing both under their
common attribute $x$, it is obvious that we have really explained
only so much of these items as {\it is x}. To explain the connection
of choke-damp and suffocation by the lack of oxygen is to leave
untouched all the other peculiarities both of choke-damp and of
suffocation---such as convulsions and agony on the one hand, density
and explosibility on the other. In a word, so far as $A$ and $B$
contain $l$, $m$, $n$, and $o$, $p$, $q$, respectively, in addition
to $x$, they are not explained by $x$. Each additional particularity
makes its distinct appeal. A single explanation of a fact only
explains it from a single point of view. The entire fact is not
accounted for until each and all of its characters have been classed
with their likes elsewhere. To apply this now to the case of the
universe, we see that the explanation of the world by molecular
movements explains it only so far as it actually {\it is\/} such
movements. To invoke the ``Unknowable'' explains only so much as is
unknowable, ``Thought'' only so much as is thought, ``God'' only so
much as is God. {\it Which\/} thought? {\it Which\/} God?---are
questions that have to be answered by bringing in again the residual
data from which the general term was abstracted. All those data that
cannot be analytically identified with the attribute invoked as
universal principle, remain as independent kinds or natures,
associated empirically with the said attribute but devoid of rational
kinship with it.

Hence the unsatisfactoriness of all our speculations. On the one
hand, so far as they retain any multiplicity in their terms, they
fail to get us out of the empirical sand-heap world; on the other, so
far as they eliminate multiplicity, the practical man despises their
empty barrenness. The most they can say is that the elements of the
world are such and such, and that each is identical with itself
wherever found; but the question Where is it found? the practical man
is left to answer by his own wit. Which, of all the essences, shall
here and now be held the essence of this concrete thing, the
fundamental philosophy never attempts to decide. We are thus led to
the conclusion that the simple classification of things is, on the
one hand, the best possible theoretic philosophy, but is, on the
other, a most miserable and inadequate substitute for the fulness of
the truth. It is a monstrous abridgment of life, which, like all
abridgments, is got by the absolute loss and casting out of real
matter. This is why so few human beings truly care for philosophy.
The particular determinations which she ignores are the real matter
exciting needs, quite as potent and authoritative as hers. What does
the moral enthusiast care for philosophical ethics? Why does the {\it
\AE sthetik\/} of every German philosopher appear to the artist an
abomination of desolation?
\bq
Grau, teurer Freund, ist alle Theorie \\
\indent Und gr\"un des Lebens goldner Baum.
\eq
The entire man, who feels all needs by turns, will take nothing as an
equivalent for life but the fulness of living itself. Since the
essences of things are as a matter of fact disseminated through the
whole extent of time and space, it is in their spread-outness and
alternation that he will enjoy them. When weary of the concrete clash
and dust and pettiness, he will refresh himself by a bath in the
eternal springs, or fortify himself by a look at the immutable
natures. But he will only be a visitor, not a dweller, in the region;
he will never carry the philosophic yoke upon his shoulders, and when
tired of the gray monotony of her problems and insipid spaciousness
of her results, will always escape gleefully into the teeming and
dramatic richness of the concrete world.
\eq

\subsubsection{The Part of Secondary Interest to the Present Discussion}

\bq
So our study turns back here to its beginning. Every way of
classifying a thing is but a way of handling it for some particular
purpose. Conceptions, ``kinds,'' are teleological instruments. No
abstract concept can be a valid substitute for a concrete reality
except with reference to a particular interest in the conceiver. The
interest of theoretic rationality, the relief of identification, is
but one of a thousand human purposes. When others rear their heads,
it must pack up its little bundle and retire till its turn recurs.
The exaggerated dignity and value that philosophers have claimed for
their solutions is thus greatly reduced. The only virtue their
theoretic conception need have is simplicity, and a simple conception
is an equivalent for the world only so far as the world is
simple---the world meanwhile, whatever simplicity it may harbor,
being also a mightily complex affair. Enough simplicity remains,
however, and enough urgency in our craving to reach it, to make the
theoretic function one of the most invincible of human impulses. The
quest of the fewest elements of things is an ideal that some will
follow, as long as there are men to think at all.

But suppose the goal attained. Suppose that at last we have a system
unified in the sense that has been explained. Our world can now be
conceived simply, and our mind enjoys the relief. Our universal
concept has made the concrete chaos rational. But now I ask, Can that
which is the ground of rationality in all else be itself properly
called rational? It would seem at first sight that it might. One is
tempted at any rate to say that, since the craving for rationality is
appeased by the identification of one thing with another, a datum
which left nothing else outstanding might quench that craving
definitively, or be rational {\it in se}. No otherness being left to
annoy us, we should sit down at peace. In other words, as the
theoretic tranquillity of the boor results from his spinning no
further considerations about his chaotic universe, so any datum
whatever (provided it were simple, clear, and ultimate) ought to
banish puzzle from the universe of the philosopher and confer peace,
inasmuch as there would then be for him absolutely no further
considerations to spin.

This in fact is what some persons think. Professor Bain says ---
\bq
\indent
A difficulty is solved, a mystery unriddled, when it can be shown to
resemble something else; to be an example of a fact already known.
Mystery is isolation, exception, or it may be apparent contradiction:
the resolution of the mystery is found in assimilation, identity,
fraternity. When all things are assimilated, so far as assimilation
can go, so far as likeness holds, there is an end to explanation;
there is an end to what the mind can do, or can intelligently desire
\ldots. The path of science as exhibited in modern ages is toward
generality, wider and wider, until we reach the highest, the widest
laws of every department of things; there explanation is finished,
mystery ends, perfect vision is gained.
\eq

But, unfortunately, this first answer will not hold. Our mind is so
wedded to the process of seeing an {\it other\/} beside every item of
its experience, that when the notion of an absolute datum is
presented to it, it goes through, its usual procedure and remains
pointing at the void beyond, as if in that lay further matter for
contemplation. In short, it spins for itself the further positive
consideration of a nonentity enveloping the being of its datum; and
as that leads nowhere, back recoils the thought toward its datum
again. But there is no natural bridge between nonentity and this
particular datum, and the thought stands oscillating to and fro,
wondering ``Why was there anything but nonentity; why just this
universal datum and not another?''\ and finds no end, in wandering
mazes lost. Indeed, Bain's words are so untrue that in reflecting men
it is just when the attempt to fuse the manifold into a single
totality has been most successful, when the conception of the
universe as a unique fact is nearest its perfection, that the craving
for further explanation, the ontological wonder-sickness, arises in
its extremest form. As Schopenhauer says, ``The uneasiness which
keeps the never-resting clock of metaphysics in motion, is the
consciousness that the non-existence of this world is just as
possible as its existence.''

The notion of nonentity may thus be called the parent of the
philosophic craving in its subtilest and profoundest sense. Absolute
existence is absolute mystery, for its relations with the nothing
remain unmediated to our understanding. One philosopher only has
pretended to throw a logical bridge over this chasm. Hegel, by trying
to show that nonentity and concrete being are linked together by a
series of identities of a synthetic kind, binds everything
conceivable into a unity, with no outlying notion to disturb the free
rotary circulation of the mind within its bounds. Since such
unchecked movement gives the feeling of rationality, he must be held,
if he has succeeded, to have eternally and absolutely quenched all
rational demands.

But for those who deem Hegel's heroic effort to have failed, nought
remains but to confess that when all things have been unified to the
supreme degree, the notion of a possible other than the actual may
still haunt our imagination and prey upon our system. The bottom of
being is left logically opaque to us, as something which we simply
come upon and find, and about which (if we wish to act) we should
pause and wonder as little as possible. The philosopher's logical
tranquillity is thus in essence no other than the boor's. They differ
only as to the point at which each refuses to let further
considerations upset the absoluteness of the data he assumes. The
boor does so immediately, and is liable at any moment to the ravages
of many kinds of doubt. The philosopher does not do so till unity has
been reached, and is warranted against the inroads of those
considerations, but only practically, not essentially, secure from
the blighting breath of the ultimate Why? If he cannot exorcize this
question, he must ignore or blink it, and, assuming the data of his
system as something given, and the gift as ultimate, simply proceed
to a life of contemplation or of action based on it.
\eq

\section{7 Jan.\ 2010 \ \ {\it What Comes Next?}\ \ \ (to M. Schlosshauer)}

Since you and I have been in so much contact recently, I asked Marcus Appleby if I could forward you the note \cite[pp.\ 1926-1927]{Samizdat2} he sent me last night.  I thought it hit the mark, and in an eloquent way:  What comes next?  That's the only really important question.

I'm feeling doubly good today.  I feel I had a very deep insight this morning just as I was getting into the shower.  It is the inevitable conclusion to draw from this QBist trek I've been taking, but I only feel that I really saw it with clarity this morning.  In a single sentence, it goes like this:  Just as quantum theory is not in conflict with probability theory, only an {\it addition\/} to it, quantum mechanics is not in conflict with the normal, empirically perceived world of common experience (sometimes called the classical world); it is only an {\it addition\/} to it.  It is not that the classical world undergirds quantum mechanics, as I read Bohr as saying; it is not that the classical world should be {\it derivable\/} from quantum mechanics (a thought that you are very familiar and probably still quite sympathetic with).  It is that one is an {\it addition\/} to the other, just as quantum theory is an addition to Bayesian probability.  It is an overpowering feeling I have in my chest, a feeling like I've only had a few times in my career---my whole body is most definitely reacting to this thought.

Anyway, I pray (metaphorically) that I can articulate this properly in some short number of paragraphs.  We shall see---I will try my best and whatever the result, it will be incorporated into the manuscript you have been reading over.  I feel I have to say it now and not wait to write another paper.

\section{4 Oct.\ 2013 \ \ {\it Script Inaccuracy} \ \ (to K. Sayenga)}

Thank you for the script, which gives me an idea of where you want to go with tomorrow's shoot. I am actually quite pleased with it.  It does quite a good job, except at the very end where the voice over says: ``Chris is one of many quantum theorists who believe there is no reality without observation.''\footnote{This refers to some planned footage for an episode of {\sl Through the Wormhole with Morgan Freeman\/}~\cite{Wormhole}.}

No, NO, {\bf NNOO}, \textbf{\textit{NNNOOOOOO!!!!!!}} You absolutely cannot say that.  I have gone out of my way my whole career to say \textit{exactly the opposite} of that.  For instance in the Schlosshauer interview here~\cite{Fuchs12} I wrote,
\bq
\ldots\ even if quantum theory is purely a theory for apportioning and structuring degrees of belief, the question of ``Why the quantum?'' is nonetheless a question of what it is about the actual, real, objective character of the world that compels us to use this framework for reasoning rather than another.
\eq
How on earth would that square with that there is no reality without observation?

My position is not at all that there is no reality without observation.  It is only that reality is not a rigid, pre-ordained, inflexible, immutable structure.  It has wiggle room, and we the observers (gambling agents) being part of reality, thus have the opportunity to participate in the universe's character and course in a nonnegligible way.

Here's a slogan:  Quantum theory concerns that part of reality that we through our actions participate in.  Quantum theory gives us the possibility that our actions upon the external world actually matter:  The world would not have been the same otherwise.  It's not that it would disappear.

You should still be able to segue into Tegmark, because his (fantasy-infused) view remains the antithesis of all that I say.  His universe really is rigid, immutable, has no room in it for our actions to be deeply meaningful in the grand scheme of things, but you can't contrast it to a kind of reality bashing on my part.

Maybe another slogan that captures my way of thinking:  The mark of reality lays in all the surprises the world gives us---the things for which we have no equations.  (For instance the unpredictable outcome of a quantum measurement.  That individual event has no equation.  Quantum mechanics gives equations for how we might gamble on it, but the event itself has no equation governing it.)  This contrasts with Tegmark's view where reality \textit{just is} a big mathematical structure---a frozen block of ice.

That would be something better for you to segue on if you want to be accurate and not mislead the public.

\section{17 Aug.\ 2014 \ \ {\it New QB Symmetries} \ \ (to S. Weinberg)}

David Mermin sent me some of his recent correspondence with you, and I am flattered that you would pay any attention to QBism at all.  However, I was not completely satisfied with David's response to you.  For one thing, I felt David presented QBism as too much of a consequence of an empiricist attitude, rather than the other way around. Myself, I would say the real story is that something akin to empiricism is \textit{forced upon us} by the quantum formalism itself.  Also, I do not feel he addressed some of your most substantial comments.  I hope I can clarify a couple of things with this note.

You pointed out, ``I don't see (and you really don't explain) how the distinction between Bayesian and other approaches to probability is relevant to this issue \ldots.''  Indeed.  Where the personalist Bayesian conception of probability comes into play is in making the original insights of Heisenberg, Pauli, and Einstein that quantum states should be viewed as ``states of knowledge'' or ``states of information'' rather than ``states of nature'' absolutely airtight in its consistency.  (Knowledge of \textit{something}, that is---in the former case measurement outcomes simpliciter, in the latter some underlying state of reality---but in both cases knowledge nonetheless.)  Elsewise, one is left with inconsistencies when the concepts are pushed hard enough:  A classic example of such an inconsistency is Bohr's insistence on the necessity of splitting the world into a quantum realm and a classical realm.  In QBism, {\it every\/} physical system is quantum in nature, and that includes the physical systems called ``humans.''  It is only by doggedly persisting with the Bayesian notion of probability in all quantum analyses that one can recover such a more-satisfying ``Copernican view'' of things, where man is not the center of the universe.

I try to argue this in the most vivid terms in this paper \cite{Fuchs10a}, which I think you would find entertaining at the least, and I believe would much help your own thinking, even if for the purposes of refining your judgment that QBism is barking up the wrong tree.

One thing I hope my analysis there makes as clear as possible is that the way one should view Bayesian probability theory is as an {\it objective\/} method for ``doing the best one can with whatever one's got.''  (The personalist nature of individual probability assignments is almost a distraction from that point of view; something not much deeper than saying that the initial condition for {\it this\/} pendulum is as ``personal'' to it as the initial condition for {\it that\/} pendulum is ``personal'' to it.  The only difference is that the initial condition is not of a position and momentum, but of a set of degrees of belief.)  And QBism says something almost identical for quantum theory:  It is an {\it objective\/} method for doing the best one can with whatever one's got when talking of quantum systems.

I liked your analogy of Darwin and Wallace.  I think it is significantly on the right track, even if we might diverge at some points.  You write, ``The great achievement of Darwin and Wallace was to show how species like humans could evolve from earlier species, without invoking any law of nature to that effect.''  Yes!  And I would say something nearly like that of quantum mechanics from the QBist perspective.  There is something about the character of the world (impersonal character, if you will) that makes quantum theory, as a methodology for decision making, ``the best we can do with what we've got.''  Or at least that's the QBist stance.  In a way, you can say that quantum theory---again as a methodology for making decisions within this (impersonal) world---has gotten locked in by selective pressures, in a way not so unlike an elephant's genome.  Both the genome and our best method for physical gambling are responses to the conditions around them.

Finally, let me say that I think all this basic conceptual work, like many of the points above, is crucial for getting oriented correctly with respect to the meaning of quantum theory (and consequently of the greatest importance in knowing how to apply it properly to fresh fields, for instance cosmology).  But far more exciting is the constructive phase of QBism:  The effort to find a new/useful/elegant formalism for quantum theory that uses probabilities only from start to finish---a formalism that never passes through the conduit of unitary and Hermitian operators and complex vector spaces.  To this end, I hope you will agree with me that Eq.\ (8) in the paper I advertised above is quite an intriguing way to rewrite the content of the Born probability rule.  It hints that we really don't know the final word on the quantum theory right here in front of us (much less of any modifications that will eventually be necessitated as physics moves forward).  But key to this new formalism is an as-yet poorly understood symmetry that seems to lie underneath all finite dimensional Hilbert spaces.  Below is an abstract from a talk I gave in March at UMass Boston that plays up exactly this point and bolsters it with some of your own words for dramatic effect.

In the end, this new symmetry may be the great brainchild of QBism for physics as a whole.  In any case, regardless of the conceptual mumbo-jumbo above, I felt you might be intrigued to learn that there appear to be hidden symmetries within the basic theory that still have not been completely uncovered even to this day.


\bq
\noindent {\bf Title:}  A New Alphabet for Quantum Information\medskip

\noindent {\bf Abstract:}  Some time ago, Steven Weinberg wrote an article for the New York Review of Books with the title, ``Symmetry:\ A `Key to Nature's Secrets'.''  So too, I would like to say of quantum information:  Only by identifying Hilbert space's most stringent and hard-to-attain symmetries will we be able to unlock quantum information's deepest secrets and greatest potential.  In this talk, I introduce the ``symmetric informationally complete'' (SIC) sets of quantum states~\cite{Fuchs2017b} as a candidate for that structure.  By their aid, one can rewrite quantum states so that they become simply probability distributions, unitary transformations so that they become doubly stochastic matrices, and the Born rule so that it becomes a rather simple variant of the classical law of total probability.  These representations hold the potential for entirely new ways of analyzing quantum communication channels and algorithms.  Paradoxically though, despite the way they can be used to make quantum theory look formally close to classical information theory, there is also a sense in which the SIC states are as far from classical as possible:  For instance, by some measures these states are as sensitive to quantum eavesdropping as any alphabet of quantum states can be.  But there is a deep question:  For all these wonderful properties, do the things actually exist?  By the end of the talk, I hope you will want them to!
\eq

\section{3 Oct.\ 2019 \ \ {\it On M \& W} \ \ (to the Reflection Team)}

I send you this note because of our recent discussions on von Fraassen's reflection principle~\cite{vanFraassen1984, Shafer1983,Goldstein1983}.  In defending our 2011 take on quantum decoherence~\cite{Fuchs2011} to Ulrich Mohrhoff, I believe I had a thought which had not come so to the front of my mind before.  Maybe R\"udiger understood it, but it strikes me only now.  It's why I'm sharing this.  Ulrich wrote,
\bum
I am well aware of you paper on decoherence. I first read it a couple of years ago, if not earlier. I completely understand your project of ``Bayesianizing'' all relevant arguments, but sometimes it reminds me (ever so faintly) of Bell's remark on efforts to get rid of the shifty split: ``It can be done --- up to a point. But it becomes embarrassing for the spectators even before it becomes uncomfortable for the snake.'' \ldots

When I first read this paper, the statement that ``Decoherence does not come conceptually before a `selection' \ldots\ but conceptually after the recognition of the future possibilities'' caused my some headache. It seems to be an example of obscuring a point by trying to make it clearer. (I find myself doing this all the time!) While every probability assignment comes after the recognition of all possible sequences of events, decoherence remains associated with a measurement (by the environment) that comes before the measurement to whose outcomes probabilities are being assigned.
\eum
To which I replied:
\bq
\noindent But the whole point of the result was to do away with talk of ``measurement by the environment'' and the contrived Hamiltonians upon which all that way of thinking requires.  A measurement in QBism is an experience of an agent full stop, and there is no place for an agent to sit in such a ``measurement by the environment'' way of thinking.  (Kind of like the little guy I made fun of in a box in Fig.\ 2 of this old paper~\cite{Caves2002}.)  Thus decoherence becomes a law of thought in QBism, like it had to.
\eq
The new thought that comes to me is that that little paper was as crucial to the consistency of QBism as was the quantum de Finetti theorem~\cite{Caves2002} years earlier.  Both in fact addressed little guys in a box that had to be exorcised for our own consistency.  Now I'm even further shamed that we did so little with that paper over the years.  We must remedy this with a fantabulous new paper.

Here's one more little piece from my replies to Ulrich that I think worth recording.  It was my claim to him that, ``I'll warn you that I am fairly skeptical of your phrase, `the incontestable irreversibility of human sensory experience.'  All one has to do is witness the decline of a grandmother or aunt with Alzheimer's disease to become skeptical of that.  At the moment, I side with Whitehead that experience too lives and dies.  Living and dying is what experience does.'' This is tied up with my (planned) response to Richard Healey over his accusation that QBism hasn't made much progress toward an ontology.  It is a question of tying up loose ends.  QBism must milk Wigner's friend for all its worth~\cite{DeBrota2020b} before it can make a leap to a Whiteheadian style ontology.  Bell asked, ``Are we not obliged to admit that more or less `measurement-like' processes are going on more or less all the time more or less everywhere?''  To that, I want to say, ``Yes, we {\it are\/} obliged! Thank you, John Bell. But measurement is not at all what you imagined it to be!''  First we have to understand what it actually is in those moments of measurement involving agents, and only then we might abstract.  To be `measurement-like' for an event is to have a subject pole and an object pole, like Whitehead's ``actual occasions,'' but to be neither intrinsically.  To be ``measurement-like'' is to be born, live, and die, like Whitehead's ``actual occasions.''  To be ``measurement-like'' is to be \ldots\ \  This takes careful thought.  This takes absolute consistency before carefully moving forward.

\section{9 Oct.\ 2020 \ \ {\it We Are Such Stuff \ldots} \ \ (to the QBuki)}

\ldots\ as quantum measurement is made on.~\cite[p.\ 27]{Fuchs10a}

Without his permission, let me quote something Marcus [Appleby] wrote me in private last week (not too big of a transgression as I had already alluded to it in our Friday meeting):
\bq
[T]o convince [Nick] Herbert and others like him we need to develop the QBist conception of reality in much more detail, and much more explicitly so that it is completely and totally obvious that we are saying something utterly different from instrumentalism.

Not that I have any clear idea how to achieve this.  But I really and deeply and strongly feel that the time has come to make an all out attack on the problem.  No more pussy-footing around the edges.  What is required is a full-frontal attack.
\eq

In this regard, let me record a quote of Emanuel Swedenborg that I just came across:  ``All things, at the present day, stand provided and prepared, and await the light.  The ship is in the harbour; the sails are swelling; the east wind blows; let us weigh the anchor, and put forth to sea.''

\section{13 Sept.\ 2020 \ \ {\it Concerning the Genuine Stuff} \ \ (to A. Gefter)} \label{GenuineStuff}

Thanks again for the notes you sent us last Friday and the cute cartoons you threw together quickly before the meeting.  I feel there's something in your heart that's very much on the right track, but we are {\it all\/} most definitely still confused.

One comment you made in your presentation caused me to wonder whether we're yet on the same page with regard to personalist Bayesianism.  It came when you showed the slide with Wigner deriving from his 50-50 assignment that it was possible that Alice would say down and he would frown.  But Alice believed that impossible.  If I understood you correctly, you were inclined to think that Alice should have the privileged judgment.  But in personalist Bayesianism, no one has a privileged judgment, even when they have a 1-0 probability assignment.

A lot of the Wignerstein paper\footnote{This refers to a lengthy treatment of Wigner's friend from a QBist perspective, tentatively titled ``QBism's Account of Wigner's Friend (and the friend's account of Wigner),'' which has become so large and patched together as to be like Frankenstein's monster.  In case that paper is never completed or posted, see  \cite{DeBrota2020b} in the meantime.}  is about establishing just that kind of idea:  That neither agent in the Wigner's friend scenario has a privileged position, either with regard to Bayesianism or quantum theory.  \v{C}aslav Brukner always likes to call Wigner a ``super-observer,'' whereas we say, ``Uh uh.  That is an error.''

In fact, it is not difficult to create examples where two agents start out with some amount of compatibility in their probability assignments, but then after one observation, completely disagree on what they will find next~\cite[fn.\ 16]{Fuchs2013}.  I.e., one will make a 1-0 probability assignment, whereas the other will make a 0-1 assignment.  Suppose they now perform the observation and the 1-0 agent gets exactly what she expected, whereas the 0-1 agent is caught completely off-guard.  What does one make of that?  Nothing more than that the first agent won the bet, whereas the second agent is probably devastated.  Whatever happened in the observation was something that one of the two equal-status agents thought would be absolutely impossible.  But even after the fact, no one has a right to claim their probability assignment was metaphysically correct beforehand.  The fact that comes about with the second observation does not establish it so.  {\it If new reality arises with each quantum measurement, then new reality arises.}

What I think this helps emphasize is a train of thought that goes all the way back to~\cite{Fuchs2002}.  When an agent writes down a POVM to describe a measurement he wishes to perform, that is equivalent to giving the potential outcomes of his soon-to-be action {\it names}.  But the actual outcomes are not equivalent to their names.  They are something much more, something much deeper than that. [This idea is significantly expanded upon in the section below titled, ``Amanda's Hooks.'']

So, here's where I definitely have sympathy for what you're aiming for.  When Alice and Bob shake hands, no matter what their expectations for the consequences of that handshake (no matter what their expectations for the experience it will elicit), no matter how they each associate a POVM with this action and its consequences, something will happen between the two of them that is very real and will go on to shape the future.  But that thing---the genuine stuff---is far more than either of the two names that Alice and Bob might name it.  The real essence of the ``connective tissue,'' the ``semi-permeable membrane'' you keep calling our attention to is something just a bit beyond what quantum theory has the resources to describe.  This is why I was calling it the pre-theoretic or pre-linguistic outcome on Friday.  It's just a name I pulled out of a hat for the occasion, but it was an attempt to express that experiences are part of the stuff of the world, but propositions like ``Alice saw the spin was up in the $x$-direction'' are but names---they are further removed from the actual stuff.\footnote{For a more formal discussion of the contents of this paragraph, see~\cite{Pienaar2021}.}

I don't yet know where all of this can go, but I do think we're starting to circle around the correct issues, and I really value that you're helping steer us that way.

Oh, let me also point you to the anti-inductivist example I told you about the other day:  See ``Priors in Quantum Bayesian Inferernce''~\cite{Fuchs2009a}.  A 2009 paper!  It seems like all the good things happened in 2009; that was the year QBism got its name~\cite{Fuchs2009a}.\footnote{Note that it was a paper appearing in one of Andrei's proceedings volumes!}

\section{3 Nov.\ 2020 \ \ {\it Consistency Is Way Overrated} \ \ (to A. Gefter)}

I don't don't have the bandwidth to be able to reply to you properly over email at the moment, so we'll just have to duke it out in person eventually.  But I do want to register that I am in disagreement with some points in your note.  Maybe the most important point for the present is this one:
\bag
To me, the interesting next question is, what happens when Wigner enters the lab and interacts with the friend? At {\it this\/} point in the story, {\it while still rejecting any 3rd person description}, there are, as far as I can tell, only two options:
\begin{enumerate}
\item
we can stick to the story that measurement outcomes are truly private and therefore Wigner and the friend can have contradictory accounts of one and the same interaction in which they both take part, or
\item
we can allow consistency to be created through the interaction itself (and because this consistency would not extend to anyone not directly interacting with Wigner and friend, it would not be 3rd person consistency, but 2nd)
\end{enumerate}
If QBism wants to stick to the first option---insisting that outcomes are truly private, and that different observers {\bf (even while interacting)} live in different spacetimes or different realities---that's fine \ldots but it is solipsistic.
\eag

I've added boldfacing to a portion of what you wrote because, even in 2012 I was already saying things like this~\cite[Ans.\ 5]{Fuchs12}:
\bq
This is where the Wigner's friend question comes into play.  This is a story of two agents with a different physical system in front of each:  1) The {\it friend}, with say an {\bf electron} in front of himself, and 2) {\it Wigner}, with the {\bf friend+electron} in front of himself.  (Agents are italicized; systems are boldfaced.)  Which agent's quantum state assignment for his own system is the correct one?  Quantum Bayesianism knows of no agent-independent notion of ``correct'' here---and this is why we say there is no paradox.  The source of each assignment is the agent who makes it, and the {\it concern\/} of each assignment is not of what is going on out in the world, but of the uncertain consequences each agent might experience if he takes any actions upon his system.  The only glaringly mutual world there is for Wigner and his friend in a QBist analysis is the partial one that might come about if these two bodies were to later take actions upon each other (``interact'')---the rest of the story is deep inside each agent's private mesh of experiences, with those having no necessary connection to anything else.

But what a limited story this is:  For its concern is only of agents and the systems they take actions upon.  What we learn from Wigner and his friend is that we all have truly private worlds in addition to our public worlds.  But QBists are not reductionists, and there are many sources of learning to take into account for a total worldview---one such comes from Nicolas Copernicus:  That man should not be the center of all things (only some things).  Thus QBism is compelled as well:  What we have learned of agents and systems ought to be projected onto all that is external to them too.  The key lesson is that each part of the universe has plenty that the rest of the universe can say {\it nothing\/} about.  That which surrounds each of us is more truly a pluriverse.
\eq

There's nothing in my writing that is as exclusionary as the way you portray me.  Particularly, I'll emphasize the last sentence of my first paragraph:  ``The only glaringly mutual world there is for Wigner and his friend in a QBist analysis is the partial one that might come about if these two bodies were to later take actions upon each other (`interact')---the rest of the story is deep inside each agent's private mesh of experiences, with those having no necessary connection to anything else.''  (This is what Cavalcanti calls the bursting of the ``Wigner bubbles'' in~\cite{Cavalcanti2021}.)

But I should have also boldfaced your phrase ``we can allow consistency to be created through the interaction itself.''  From my point of view, consistency is way overrated:  Agents need not have agreement to live in the same world.  {\it The very fact that they are fighting it out (perhaps to the death) is already indication enough that they are in a single world.}  Just look at the Democrats and the Trumpists on this very day.  What one declares to be true, the other declares to be a lie:  They fight it out precisely because they are in one world.  In fact, for the Darwin-inspired ontology the classical pragmatists wanted, they nor I would want it any other way:  This is one of the mechanisms by which novelty seeps into the pluriverse.

So when you want to find a yet-unearthed mechanism within QBism for how consistency between Wigner and the friend is {\it enforced\/} or {\it necessitated}, I think you are on the wrong track.  Wigner and the friend {\it may\/} or {\it may not\/} agree when they eventually speak to each other about the electron:  It is not within the {\it raw\/} quantum formalism (the user's guide for individual agents) to say.  What Rovelli and everyone previously have left out of their analyses is that the problem is simply way underdetermined.  Before our recent paper with R\"udiger~\cite{DeBrota2020b}, I don't believe anyone had ever given a flip about treating Wigner and the friend on the same footing, completely symmetrically.  But then once one does that, one sees that Wigner and his friend can disagree just as any two Bayesians can.

This is not at all to say that one could not identify various conditions by which to adorn the friend's and Wigner's uses of the quantum formalism so that they feel confident they will come to agreement upon speaking to each other.  But then such an addition would be in the spirit of, for instance, de Finetti's assumption of exchangeability for his representation theorem.  I.e., he identifies a condition (a smidgen of agreement they have beforehand) through which two agents believe they will come to complete agreement after viewing enough data.  The point I want to make is this:  The belief in an ultimate common agreement is not bought from the {\it raw\/} formalism of probability theory, but from a certain feature common to the two Bayesians' otherwise distinct priors.  The formalism by itself cannot set the priors.  QBists expect it to be no different when it comes to quantum theory---in fact, QBism is pretty much built on that very idea!  In fact, Jacques has spent a little time trying to go further with the Wigner's friend scenario, trying to find some extra conditions that would lead to the kind of assurances you want, but I believe he'll tell you that it's not immediately obvious how to proceed.

But I said I don't have so much bandwidth at the moment.  There's still a lot more to talk about!  So, let me just shut down the note, and in the spirit of Michael Murphy, say ``To be continued!''

\section{12 Feb.\ 2021 \ \ {\it Getting Merleau-Pontified} \ \ (to J. Pienaar)}

Jacques, I felt it a very productive and stimulating discussion today. Thanks again! I am genuinely starting to get Merleau-Pontified by you.

I loved the line of yours about how in M-P's hands phenomenology becomes not an epistemology, but an ontology.  It reminded me of this little bit of William James that I always come back to:
\bq
The import of the difference between pragmatism and rationalism is now in sight throughout its whole extent.  The essential contrast is that {\it for rationalism reality is ready-made and complete from all eternity, while for pragmatism it is still in the making, and awaits part of  its complexion from the future}.  On the one side the universe is absolutely secure, on the other it is still pursuing its adventures.  [\ldots]

{\it The alternative between pragmatism and rationalism, in the shape in which we now have it before us, is no longer a question in the theory of knowledge, it concerns the structure of the universe itself.}

On the pragmatist side we have only one edition of the universe, unfinished, growing in all sorts of places, especially in the places where thinking beings are at work.

On the rationalist side we have a universe in many editions, one real one, the infinite folio, or {\it \'edition de luxe}, eternally complete; and then the various finite editions, full of false readings, distorted and mutilated each in its own way.
\eq
It's not about a theory of knowledge, but about the structure of the universe itself.

By the way, on my bookshelf I have a small book titled {\sl William James and the Reinstatement of the Vague\/} by William Joseph Gavin, which I skimmed probably 10 years ago.  On the back cover, a reviewer writes, ``Gavin has taken an often cited but seldom explored text from James and shown its significance for James's overall philosophy.  He clearly and insightfully delineates the character and role of `the vague' in James's metaphysics \ldots'' \ On page 1, Gavin writes:
\bq
In {\sl The Principles of Psychology}, the American philosopher William James makes what at first sight appears as a strange statement.  He tells the reader:  ``It is, in short, the re-instatement of the vague to its proper place in our mental life which I am so anxious to press on the attention.''  \ldots\  In {\sl Pragmatism}, he tells the reader, ``Profusion, not economy, may after all be reality's key-note''; and asks: ``May there not after all be a possible ambiguity in truth?''  And in {\sl Essays in Radical Empiricism}, he calls ``our experiences, taken all together, a quasi-chaos.''  Now most philosophers, and indeed most people, do not set out to emphasize the vague.  Quite the opposite:  most have pursued certainty, objectivity, and some form of universal truth.  Furthermore, most have assumed that such truth can be captured in language, that is, that a complete linguistic description of reality is at least possible, and certainly desirable.  Yet James does not.
\eq
It might be useful to do a comparative study between M-P's ``partiality'' and James's ``the vague.''

Finally, here are a couple of passages from my paper~\cite{Fuchs2017} which I tried unsuccessfully to recall today:
\bq
There is a sense in which this unhinging of the Born Rule from being a ``law of nature'' in the usual conception---i.e., treating it as a normative statement, rather than a descriptive one---makes the QBist notion of quantum {\it indeterminism\/} a far more radical variety than anything proposed in the quantum debate before.  It says in the end that nature does what it wants, without a mechanism underneath, and without any ``hidden hand'' of the likes of Richard von Mises's {\it Kollective\/} or Karl Popper's {\it propensities\/} or David Lewis's {\it objective chances\/}, or indeed any conception that would diminish the autonomy of nature's events.  Nature and its parts do what they want, and we as free-willed agents do what we can get away with.  Quantum theory, on this account, is our best means yet for hitching a ride with the universe's pervasive creativity and doing what we can to contribute to it.
\eq

\bq
The case QBism makes before the forum is this.  What the quantum agent---the protagonist in the drama of any application of quantum theory---is ultimately doing is hitching a ride with a {\it new kind of ontology or metaphysic}.  An ontology of all-pervasive, {\it pan-creative}\footnote{This terminology is adapted from, but may not be identical to, Michel Weber's designation of Alfred North Whitehead's later ontology as a {\it pancreativism}~\cite{Weber2006,Weber2011}.} experience, not unakin to what William James was thinking of when he wrote his essay ``Does `Consciousness' Exist?''
\eq

To the extent that I understand it, I think M-P did well to make a distinction between ``perception'' and ``phenomenon,'' of which my present understanding is this:  The first refers to a particular person's experience, while the second attempts to depersonalize the concept and apply it more widely.  To mix metaphors, ``perception'' refers to those ``drops of experience'' to which an agent is hitching a ride, and ``phenomena'' refer to those that might be outside any human's outfit.  Not making such a linguistic distinction is potentially one of the things that helped derail James and Dewey's program of an ontology of ``pure experience.''  Here's a passage from Martin Jay's book {\sl Songs of Experience: Modern American and European Variations of a Universal Theme\/}:
\bq
Was it ever really possible, other critics asked, to suppress the subjective connotation of the very word ``experience,'' as both James and Dewey insisted was necessary, and allow it to stand for something deeper than the subject/object split?  Could experience, moreover, serve as a foundationless alternative to traditional grounding concepts in philosophy, one based on accepting uncertainty rather than seeking to overcome it, or was it merely another variation of the same fruitless quest?  Was it, in the final analysis, really anything more than what one critic damned as ``a vague, incantatory expression of Dewey's,'' which served less to answer hard questions than postpone them?

That Dewey himself came to realize he was fighting an uphill battle is evident in a frequently quoted admission he made in a revised edition of {\sl Experience and Nature}, where he wrote in frustration: ``I would abandon the term `experience' because of my growing realization that the historical obstacles which prevented understanding of my use of `experience' are, for all practical purposes, insurmountable.''  In its place, he mused, he would now want to put the word ``culture,'' thus duplicating virtually the same move made by Raymond Williams a generation later (and which would have surely led into the same swamps of meaning that almost drowned his British counterpart).  In his final work, jointly written with Arthur Bentley, {\sl Knowing and the Known}, Dewey ruefully acknowledged that ``experience'' would no longer do the work he had thought it might, preferring ``trans-action'' instead, even if the basic premises of his thought had not really changed.
\eq

Anyway, again, a really stimulating discussion for me today.  I'm looking forward to next week's.

\section{3 Mar.\ 2021 \ \ {\it Amanda's Hooks} \ \ (to the QBuki)}

It is hard for me keep up with all the responses {\it I dream of making\/} to so many of your notes.  I dream of writing and writing and all that happens is that Zoom keeps interrupting my time instead.  A case in point was the note Amanda wrote to all of us on February 26:
\bag
Let's say, on the one hand, I take for granted all the levels of personalism that Blake was mentioning. But let's say I also take seriously Merleau-Ponty's ``reversibility of the flesh'' and the notion of a shared interaction, which says that when I interact directly with another agent (I'm Wigner asking my friend what outcome she got after breaking the hermetic seal of the lab), the outcome of my measurement of my friend can also be read in reverse as my friend's outcome of her measurement on me. Can I then say, that after my measurement of my friend, with its outcome personal to me, I update my beliefs/priors, which will then inform future quantum state assignments I give to relevant measurements---and meanwhile my friend updates her beliefs/priors, which will then inform future quantum state assignments she makes on relevant measurements, and that while all of that is happening in our own ``island universes,'' the reversibility of our original interaction means that we should be updating our priors ``in the same direction,'' as it were, and therefore that while our quantum state assignments will never be the same, because we will always have different contexts as a result of our having different bodies with different histories, they can at least begin to converge as a result of our interactions? Or have I violated QBism in some way in that scenario?
\eag

This caused a great excitement in me, and I jotted down very quickly:
\bq
\noindent There are some elements I really like in this!!  I don't think it is convergence necessarily (though it could be), but a kind of synchronic update (the consequence for each of them) for each agent.  A natural mapping between consequences.  \ldots\ I think there is a road to progress in this thought!
\eq
and then a little bit later
\bq
\noindent I should have said ``between their individual consequences.''  In Merleau-Ponty speak, these considerations concern all that happens after the reflective individuation occurs.  I.e.\ the level at which the quantum formalism might be made use of.  At lower levels of reality, there's no good sense to it.  At best it can inform us of the lower (actually {\it wider\/}) levels of reality.
\eq
{\it But\/} I never {\it adequately\/} responded to express all the flurry of thoughts Amanda provoked in me.

So treating myself as in an Indiana Jones movie, leaping through an exit just as a great stone door is flying toward the ground and reaching back for the dropped crystal in the nick of time to still save my arm \ldots\  Let me try only to say the essential words and add in a couple of previous notes to Amanda to do a little explanatory work.

The epiphany in me (or more accurately the d\'ej\`a vu) provoked by Amanda's remarks was this idea:  The formal expression that Wigner and his friend in their handshake are two poles of the same ``genuine stuff'' (as I called it in the section titled ``Concerning the Genuine Stuff'' above) is that Wigner and friend both use the apparatus of quantum to update from prior to posterior states.  The handshake is the ``hook'' or ``pivot'' or ``spur'' upon which they base an update.  This alleviates the problem that intersubjective agreement must necessarily find a place either in the individual practices or experimental consequences of quantum theory.

When I wrote to Amanda in the section titled ``Consistency Is Way Overrated,'' ``From my point of view, consistency is way overrated:  Agents need not have agreement to live in the same world.  The very fact that they are fighting it out (perhaps to the death) is already indication enough that they are in a single world,'' I didn't have quite this clarity in my mind.  The key point I want to try to express is that I have always been suspicious of intersubjectivity as a criterion for agents to be said to be living in a common world, and I think Amanda's remarks got me a little closer to clarifying what the real issue is here.

Now, why did I say ``more accurately the d\'ej\`a vu”?  In the abstract to my monstrous samizdat~\cite{Fuchs2014}, I wrote, ``More roundly, the document is an attempt to provide an essential ingredient, unavailable anywhere else, for turning QBism into a live option within the vast spectrum of quantum foundational thought.''  How I should heed my own words and look through it from time to time.  Take this passage from 4 November 2002, essentially 18 years before I had ever heard of Merleau-Ponty:
\bq
We generally write a POVM as an indexed set of operators, $E_d$. Here
is how I would denote the referents of those symbols.  The index $d$
should be taken to stand for the raw data that can enter our
attention when a quantum measurement is performed.  The whole object
$E_d$ should be construed as the ``meaning'' we propose to ascribe to
that piece of data when/if it comes to our attention.  It is
important here to recognize the logical distinction between these two
roles.  The symbol $d$ stands for something beyond our control,
something that enters into us from the world outside our head.  The
ascription of a particular value $d$ is not up to us, by definition.
The {\it function\/} $E_d$, however, is of a completely different
flavor. It is set by our history, by our education, by whatever
incidental factors that have led us to believe whatever it is that we
believe when we walk into the laboratory to elicit some data.  That
is to say, $E_d$ has much the character of a subjective probability
assignment.  It is a judgment. [\ldots]

[With that in mind,] here's a point of view I'm finding myself more and more attracted to lately.

I think it is safe to say that the following idea is pretty
commonplace in quantum mechanical practice.  Suppose I measure a
single POVM twice---maybe on the same system or two different
systems, I don't care---and just happen to get the same outcome in
both cases.  Namely, a single operator $E_d$.  The common idea, and
one I've held onto for years, is that there is an objective sense in
which those two events are identical copies of each other.  They are
like identical atoms \ldots\ or something like the spacetime
equivalent of atoms.  But now I think we have no warrant to think
that.  Rather, I would say the two outcomes are identical only
because we have (subjectively) chosen to ignore almost all of their
structure.

That is to say, I now count myself not so far from the opinion of
Ulfbeck and Bohr, when they write~\cite{Ulfbeck2001}:
\bq
\noindent
   The click \ldots\ is seen to be an event entirely beyond law.  \ldots\ [I]t
   is a unique event that never repeats \ldots\ The uniqueness of the click,
   as an integral part of genuine fortuitousness, refers to the click in
   its entirety \ldots . [T]he very occurrence of laws governing the clicks
   is contingent on a lowered resolution.
\eq
For though I have made a logical distinction between the role of the
$d$'s and the $E_d$'s above, one should not forget the very
theory-ladenness of the set of possible $d$'s.  What I think is going
on here is that it takes (a lot of) theory to get us to even
recognize the raw data, much less ascribe it some meaning.  In Marcus
Appleby's terms, all that stuff resides in the ``primitive theory''
(or perhaps some extension of it), which is a level well below
quantum mechanics.  What quantum mechanics is about is a little froth
on the top of a much deeper sea.  Once that deeper sea is set, then
it makes sense to make a distinction between the inside and the
outside of the agent---i.e., the subjective and the objective---as we
did above. For even in this froth on the top of a deeper sea, we
still find things we cannot control once our basic beliefs---i.e.,
our theory---are set.

Without the potential $d$'s we could not even speak of the
possibility of experiment.  Yet like the cardinality of the set of
colors in the rainbow---Newton said seven, Aristotle said three or
four---a subjective judgment had to be made (within the wide
community) before we could get to that level.  If this is so, then it
should not strike us as so strange that the raw data $d$ in our
quantum mechanical experience will ultimately be ascribed with a
meaning $E_d$ that is subjectively given.  (I expressed some of this
a little better in a note I wrote to David last month; I'll place it
below as a supplement.)  More particularly, with respect to the EPR
example above, it should not strike us as odd that the phenomenon
comes about solely because of an interpretive convention we set:  All
quantum measurement outcomes are unique and incomparable at the ontic
level.  At least that's the idea I'm toying with.
\eq

In a later note from 28 July 2003, I complained to my colleague R\"udiger that he didn't have enough ontological inclinations for my tastes:
\bq
In your newest turn, such discussion would be ``meaningless''---correct?  You would no longer allow yourself to contemplate what the {\it click\/} $d$ might be in its own essence.  You would no longer say that ``successive clicks with a single value $d$ are truly the same,'' nor would you contemplate that ``successive clicks with a single value $d$ may actually be truly different.''

That is to say, in this new philosophy you are toying with, if something (i.e., an event, a fact, etc.) is not a ``hook'' upon which a probability can be conditioned, you are not willing to speak of it at all.  It is meaningless---I believe you say.  You don't even let yourself conjecture about the stuff that is out there independently of us. (Or at least I don't see how you're going to be able to do this with such a strongly positivistic line.)

As I tried to express today, and as I'm trying harder to articulate now, I guess I don't like that.  Pragmatists are not positivists, but more opportunists.
\eq

Attached is the full discussion for anyone who cares to look.  Also attached is a nice find by Jacques today that's quite relevant to this very discussion~\cite{Nielsen2020}.  Its abstract reads, ``For two ideally rational agents, does learning a finite amount of shared evidence necessitate agreement?  No.  But does it at least guard against belief polarization, the case in which their opinions get further apart?  No.  OK, but are rational agents guaranteed to avoid polarization if they have access to an infinite, increasing stream of shared evidence?  No.''

OK, time for QBuki.  One more articulated response down; 458 to go!

\section{Acknowledgements}

I thank Blake Stacey for a number of suggestions about the material included herein and Arkady Plotnitsky and Emmanuel Haven for infinite patience.  This work was supported in part by the John E. Fetzer Memorial Trust.


\begin{thebibliography}{99}

\bibitem{FuchsStacey2018}
C. A. Fuchs and B. C. Stacey, ``QBism:\ Quantum Theory as a Hero's Handbook,'' in {\sl Proceedings of the International School of Physics ``Enrico Fermi'' Course 197 -- Foundations of Quantum Physics}, edited by E.~M. Rasel, W.~P. Schleich, and S. W\"olk (IOS Press, Amsterdam; Societ\`a Italiana di Fisica, Bologna, 2018), pp.\ 133--202; \href{https://arxiv.org/abs/1612.07308}{\tt arXiv:1612.07308}.

\bibitem{Fuchs2017}
C. A. Fuchs, ``Notwithstanding Bohr, the Reasons for QBism,'' {\sl Mind and Matter\/} {\bf 15}, 245--300 (2017);  \href{https://arxiv.org/abs/1705.03483}{\tt arXiv:1705.03483}.


\bibitem{Fuchs2016}
C. A. Fuchs, ``On Participatory Realism,'' in {\sl Information and Interaction:\ Eddington, Wheeler, and the Limits of Knowledge}, edited by I.~T. Durham and D. Rickles, (Springer, Berlin, 2016), pp.\ 113--134; \href{https://arxiv.org/abs/1601.04360}{\tt arXiv:1601.04360}.

\bibitem{CFS02}
C. M. Caves, C. A. Fuchs and R. Schack, ``Quantum Probabilities as Bayesian Probabilities,'' {\sl Physical Review A\/} {\bf 65}, 022305 (2002).

\bibitem{Fuchs2001}
C. A. Fuchs, ``Quantum Foundations in the Light of Quantum
Information,'' in {\sl Decoherence and its Implications in Quantum
Computation and Information Transfer:\ Proceedings of the NATO
Advanced Research Workshop, Mykonos Greece, June 25--30, 2000},
edited by A.~Gonis and P.~E.~A. Turchi (IOS Press, Amsterdam, 2001),
pp.\ 38--82; \href{https://arxiv.org/abs/quant-ph/0106166}{\tt quant-ph/0106166}.

\bibitem{Timpson08a}
C. G. Timpson, ``Quantum Bayesianism:\ A Study,'' {\sl Studies in History and Philosophy of Modern Physics}\ {\bf 39}, 579--609 (2008); \href{https://arxiv.org/abs/0804.2047}{\tt arXiv:0804.2047}.

\bibitem{Jaynes2003}
E. T. Jaynes, {\sl Probability Theory:\ The Logic of Science}, (Cambridge University Press, Cambridge, UK, 2003).

\bibitem{deFinetti1990}
B. de Finetti, {\sl Theory of Probability}, (Wiley, New York, 1990).

\bibitem{Fuchs2002}
C. A. Fuchs, ``Quantum Mechanics as Quantum Information (and only a little more),'' in {\sl Quantum Theory:\ Reconsideration of
Foundations}, edited by A.~Khrennikov ({\Vaxjo} University Press, {\Vaxjo}, Sweden, 2002), pp.\ 463--543; \href{https://arxiv.org/abs/quant-ph/0205039}{\tt arXiv:quant-ph/0205039}.

\bibitem{Fuchs10a}
C.~A. Fuchs, ``QBism, the Perimeter of Quantum Bayesianism,'' \href{https://arxiv.org/abs/1003.5209}{\tt arXiv:1003.5209}.

\bibitem{Stacey2019a}
B. C. Stacey, ``Ideas Abandoned en Route to QBism,'' \href{https://arxiv.org/abs/1911.07386}{\tt arXiv:1911.07386}.

\bibitem{Samizdat2}
C.~A. Fuchs, {\sl My Struggles with the Block Universe:\ Selected Correspondence, January 2001 -- May 2011}, edited by Blake C. Stacey, foreword by Maximilian Schlosshauer (2014), 2,349 pages; \href{https://arxiv.org/abs/1405.2390}{\tt arXiv:1405.2390}.

\bibitem{FuchsBreak}
C. A. Fuchs, {\sl Quantum States:\ What the Hell Are They?\ The Post-{\Vaxjo} Phase Transition} (2002). [The document should soon be available at \href{https://qbism.org}{\tt https://qbism.org}.]

\bibitem{NPI1}
C.~A. Fuchs, {\sl Notes on a Paulian Idea:\ Foundational, Historical, Anecdotal \& Forward-Looking Thoughts on the Quantum},
with foreword by N. David Mermin, 504 pages; \href{https://arxiv.org/abs/quant-ph/0105039}{\tt arXiv:quant-ph/0105039}.

\bibitem{NPI2}
C.~A. Fuchs, {\sl Notes on a Paulian Idea:\ Foundational, Historical, Anecdotal \& Forward-Looking Thoughts on the Quantum},
with foreword by N. David Mermin, ({\Vaxjo} University Press, {\Vaxjo}, Sweden, 2003), 718 pages.

\bibitem{Fuchs2010}
C.~A. Fuchs, \href{https://www.amazon.com/Coming-Age-Quantum-Information-Paulian/dp/0521199263}{\sl Coming of Age with Quantum Information:\ Notes on a Paulian Idea}, (Cambridge University Press, Cambridge, UK, 2010), 598 pages.

\bibitem{Greenberger2011}
D. M. Greenberger, book review, ``{\sl Coming of Age with Quantum Information:\ Notes on a Paulian Idea}.\ Christopher A. Fuchs,'' {\sl American Journal of Physics\/} {\bf 79}, 1083--1084 (October 2011).

\bibitem{Cavalcanti2011}
E. Cavalcanti, ``\href{https://www.proquest.com/docview/916790900}{Quantum Subversives},'' {\sl American Scientist\/} {\bf 99}, 500--502 (2011).

\bibitem{Trabesinger2011}
A. Trabesinger, ``\href{https://www.nature.com/articles/nphys2020}{Inside Quantum Information},'' {\sl Nature Physics\/} {\bf 7}, 443--444 (June 2011).

\bibitem{Benedictus2012}
F. Benedictus, ``\href{https://link.springer.com/article/10.1007/s11016-011-9637-y}{Quantum Information},'' {\sl Metascience\/} {\bf 21}, 595--600 (November 2012).

\bibitem{Cavalcanti2012}
E.~Cavalcanti, book review, ``Christopher A. Fuchs:\ {\sl Coming of age with quantum information:\ notes on a Paulian idea}, {\sl Quantum Information Processing\/} {\bf 11}, 633--636 (April 2012).

\bibitem{Dick2013}
S. Dick, book review, ``Christopher A. Fuchs.\ {\sl Coming of Age with Quantum Information:\ Notes on a Paulian Idea},'' {\sl Isis\/} {\bf 104}, 646--647 (September 2013).

\bibitem{Morikoshi2015}
F. Morikoshi, book review, ``C. A. Fuchs, {\sl Coming of Age with Quantum Information:\ Notes on a Paulian Idea},'' {\sl Nihon Butsuri Gakkaishi} {\bf 70}(3), 224 (2015), in Japanese.

\bibitem{Bilban2020}
T. Bilban, book review, ``Christopher A. Fuchs:\ Coming of Age with Quantum Information,'' {\sl Phainomena\/} {\bf 29}(114/115), 223--229, 1 December 2020, in Slovenian.

\bibitem{Fuchs2021}
C. A. Fuchs, ``\href{https://doi.org/10.1007/s11007-020-09525-6}{Interview with Physicist Christopher Fuchs},'' with Robert P. Crease and James Sares, {\sl Continental Philosophy Review}, (5 February 2021).

\bibitem{Wiltsche2020}
H. A. Wiltsche and P. Berghofer, ``Phenomenological Approaches to Physics:\ Mapping the Field,'' in {\sl Phenomenological Approaches to Physics}, Synthese Library {\bf 429}, edited by H. A. Wiltsche and P. Berghofer (Springer, Cham, Switzerland, 2020), pp.\ 1--47.

\bibitem{Bitbol2020a}
M. Bitbol, ``\href{http://philsci-archive.pitt.edu/19512/}{A Phenomenological Ontology for Physics:\ Merleau-Ponty and QBism},'' in {\sl Phenomenological Approaches to Physics}, Synthese Library {\bf 429}, edited by H. A. Wiltsche and P. Berghofer (Springer, Cham, Switzerland, 2020), pp.\ 227--242.

\bibitem{Tremblaye2020}
L. de La Tremblaye, ``QBism from a Phenomenological Point of View:\ Husserl and QBism,'' in {\sl Phenomenological Approaches to Physics}, Synthese Library {\bf 429}, edited by H. A. Wiltsche and P. Berghofer (Springer, Cham, Switzerland, 2020), pp.\ 243--260.

\bibitem{Bitbol2020b}
M. Bitbol, ``\href{https://doi.org/10.1007/s11007-020-09515-8}{Is the Life-World Reduction Sufficient in Quantum Physics?},'' {\sl Continental Philosophy Review}, (17 October 2020).

\bibitem{Berghofer2020}
P. Berghofer, ``\href{https://link.springer.com/article/10.1007/s13194-020-00294-w}{Scientific Perspectivism in the Phenomenological Tradition},'' {\sl European Journal for Philosophy of Science\/} {\bf 10}, 30 (2020).

\bibitem{Boge2021}
F. J. Boge, ``Realism without Interphenomena:\ Reichenbach's Cube, Sober's Evidential Realism, and Quantum Solipsism,'' {\sl International Studies in the Philosophy of Science}, (14 August 2021).

\bibitem{Pienaar2022}
J. Pienaar, ``Unobservable Entities in QBism and Phenomenology,'' forthcoming in {\sl Phenomenology and QBism:\ New Approaches to Quantum Mechanics}, edited by P.~Berghofer and H.~A.~Wiltsche (Routledge, New York, 2022).

\bibitem{Boge2022}
F. J. Boge, ``Back to Kant!\ QBism, Phenomenology, and Reality from Invariants,'' forthcoming in {\sl Phenomenology and QBism:\ New Approaches to Quantum Mechanics}, edited by P.~Berghofer and H.~A.~Wiltsche (Routledge, New York, 2022).

\bibitem{French2022}
S. French, ``\href{http://philsci-archive.pitt.edu/19354/}{Putting Some Flesh on the Participant in Participatory Realism},'' forthcoming in {\sl Phenomenology and QBism:\ New Approaches to Quantum Mechanics}, edited by P.~Berghofer and H.~A.~Wiltsche (Routledge, New York, 2022).

\bibitem{Bitbol2022}
M. Bitbol and L. de La Tremblaye, ``QBism:\ An Eco-Phenomenology of Quantum Physics,'' forthcoming in {\sl Phenomenology and QBism:\ New Approaches to Quantum Mechanics}, edited by P.~Berghofer and H.~A.~Wiltsche (Routledge, New York, 2022).

\bibitem{Baeyer2022}
H. C. von Baeyer, ``On the Consilience between QBism and Phenomenology,'' forthcoming in {\sl Phenomenology and QBism:\ New Approaches to Quantum Mechanics}, edited by P.~Berghofer and H.~A.~Wiltsche (Routledge, New York, 2022).

\bibitem{Fuchs02b}
C.~A. Fuchs, ``The Anti-{\Vaxjo} Interpretation of Quantum
Mechanics,'' in {\sl Quantum Theory:\ Reconsideration of Foundations},
edited by A.~Khrennikov ({\Vaxjo} University Press, {\Vaxjo}, Sweden, 2002), pp.\ 99--116; \href{https://arxiv.org/abs/quant-ph/0204146}{\tt arXiv:quant-ph/0204146}.

\bibitem{Fuchs07a}
C.~A. Fuchs, ``Delirium Quantum: Or, where I will take quantum mechanics if it will let me,'' in {\sl Foundations of Probability and Physics -- 4}, edited by G.~Adenier, C.~A. Fuchs, and A.~Yu.\ Khrennikov, AIP Conference Proceedings Vol.~889, (American Institute of Physics, Melville, NY, 2007), pp.\ 438--462; \href{https://arxiv.org/abs/0906.1968}{\tt arXiv:0906.1968}.

\bibitem{Fuchs2014}
C. A. Fuchs and R. Schack, ``Quantum Measurement and the Paulian Idea,'' in {\sl The Pauli-Jung Conjecture and Its Impact Today}, edited by H.~Atmanspacher and C.~A. Fuchs (Imprint Academic, Exeter, UK, 2014), pp.\ 93--107; \href{https://arxiv.org/abs/1412.4209}{\tt arXiv:1412.4209}.

\bibitem{Pienaar2020}
J. Pienaar, ``Extending the Agent in QBism,'' {\sl Foundations of Physics\/} {\bf 50}, 1894--1920 (2020); \href{https://arxiv.org/abs/2004.14847}{\tt arXiv:2004.14847}.

\bibitem{Pienaar2021}
J. Pienaar, ``QBism and Relational Quantum Mechanics Compared,'' \href{https://arxiv.org/abs/2108.13977}{\tt arXiv:2108.13977}.

\bibitem{Schroedinger54}
E. Schr\"odinger, {\sl Nature and the Greeks and Science and Humanism}, (Cambridge University Press, Cambridge, UK, 2014).

\bibitem{Moller2008}
Mark Moller, ``\,`The Many and the One' and the Problem of Two Minds Perceiving the Same Thing,'' {\sl William James Studies\/} {\bf 3}, (2008).

\bibitem{Peres2005}
A. Peres, ``Einstein, Podolsky, Rosen, and Shannon,'' {\sl Foundations of Physics\/} {\bf 35}, 511--514 (2005); \href{https://arxiv.org/abs/quant-ph/0310010}{\tt arXiv:quant-ph/0310010}.

\bibitem{DeBrota2021}
J. B. DeBrota, C. A. Fuchs, J. L. Pienaar, and B. C. Stacey, ``Born's Rule as a Quantum Extension of Bayesian Coherence,'' {\sl Physical Review A\/} {\bf 104}, 022207 (2021).

\bibitem{DeBrota2020}
J. B. DeBrota, C. A. Fuchs, and B. C. Stacey, ``Symmetric Informationally Complete Measurements Identify the Irreducible Difference between Classical and Quantum Systems,'' {\sl Physical Review Research\/} {\bf 2}, 013074 (2020).

\bibitem{Appleby2017}
D. M. Appleby, C. A. Fuchs, B. C. Stacey, and H. Zhu, ``Introducing the Qplex:\ A Novel Arena for Quantum Theory,'' {\sl The European Physical Journal D\/} {\bf 71}, 197 (2017).

\bibitem{Appleby2015}
D. M. Appleby, C. A. Fuchs, and H. Zhu, ``Group Theoretic, Lie Algebraic and Jordan Algebraic Formulations of the SIC Existence Problem,'' {\sl Quantum Information and Computation\/} {\bf 15}, 61--94 (2015); \href{https://arxiv.org/abs/1312.0555}{arXiv:1312.0555}.

\bibitem{PD2009}
Y.~Aharonov, C.~A. Fuchs, S.~Goldstein, T.~Norsen, T.~Rudolph, R.~W. Spekkens, and C.~G. Timpson, ``PIAF 09 -- Panel Discussion,'' 30 September 2009, \href{https://pirsa.org/09090098}{\tt PIRSA:09090098}.

\bibitem{Mermin2009}
N. D. Mermin, ``Confusing Ontic and Epistemic Causes Trouble in Classical Physics Too,'' talk at PIAF 09 New Perspectives on the Quantum State, 27 September 2009, \href{https://pirsa.org/09090077}{PIRSA:09090077}.

\bibitem{vonBaeyer2009}
H. C. von Baeyer, {\sl Petite Le\c{c}ons de Physique dans les Jardins de Paris}, (Dunod, Paris, 2009).

\bibitem{Wormhole}
C. A. Fuchs, segment devoted to QBism in episode titled ``Is Luck Real?''\ of the Science Channel's television show {\sl Through the Wormhole with Morgan Freeman}, March 2014; video clip posted at \href{https://www.youtube.com/watch?v=LQvCTZgNRNw}{\tt https://youtu.be/LQvCTZgNRNw}.

\bibitem{Fuchs12}
C. A. Fuchs, ``Interview with a Quantum Bayesian,'' in {\sl Elegance and Enigma:\ The Quantum Interviews}, edited by M.~Schlosshauer (Springer, Berlin, Frontiers Collection, 2011); \href{https://arxiv.org/abs/1207.2141}{\tt arXiv:1207.2141}.

\bibitem{Fuchs2017b}
C. A. Fuchs, M. C. Hoang, and B. C. Stacey, ``The SIC Question:\ History and State of Play,'' {\sl Axioms\/} {\bf 6}, 21 (2017); \href{https://arxiv.org/abs/1703.07901}{\tt arXiv:1703.07901}.

\bibitem{vanFraassen1984}
B. C. van Fraassen, ``Belief and the Will,''  {\sl Journal of Philosophy\/} {\bf 81}, 235--256 (1984).

\bibitem{Shafer1983}
G. Shafer, ``A Subjective Interpretation of Conditional Probability,'' {\sl Journal of Philosophical Logic\/} {\bf 12}, 453--466 (1983).

\bibitem{Goldstein1983}
M. Goldstein, ``The Prevision of a Prevision,'' {\sl Journal of the American Statistical Association\/} {\bf 78}, 817--819 (1983).

\bibitem{Fuchs2011}
C.~A. Fuchs and R.~Schack, ``Bayesian Conditioning, the Reflection Principle, and Quantum Decoherence,'' in {\sl Probability in Physics}, edited by Y.~Ben-Menahem and M.~Hemmo (Springer, Berlin, Frontiers Collection, 2012), pp.\ 233--247; \href{https://arxiv.org/abs/1103.5950}{\tt arXiv:1103.5950}.

\bibitem{Caves2002}
C.~M. Caves, C.~A. Fuchs and R.~Schack, ``Unknown Quantum States:\ The Quantum de Finetti Representation,'' {\sl Journal of Mathematical Physics\/} {\bf 43}, 4537--4559 (2002); \href{https://arxiv.org/abs/quant-ph/0104088}{\tt arXiv:quant-ph/0104088}.

\bibitem{DeBrota2020b}
J. B. DeBrota, C. A. Fuchs, and R. Schack, ``Respecting One's Fellow:\ QBism's Analysis of Wigner's Friend,'' {\sl Foundations of Physics\/} {\bf 50}, 1859--1874 (2020); \href{https://arxiv.org/abs/2008.03572}{\tt arXiv:2008.03572}.

\bibitem{Fuchs2009a}
C.~A. Fuchs and R.~Schack, ``Priors in Quantum Bayesian Inference,'' in {\sl Foundations of Probability and Physics -- 5}, edited by L.~Accardi et al., AIP Conference Proceedings Vol.~1101, (American Institute of Physics, Melville, NY, 2009), pp.\ 255--259; \href{https://arxiv.org/pdf/0906.1714.pdf}{\tt arXiv:0906.1714}.

\bibitem{Fuchs2009b}
C.~A. Fuchs and R.~Schack, ``From Quantum Interference to Bayesian Coherence and Back Round Again,'' in in {\sl Foundations of Probability and Physics -- 5}, edited by L.~Accardi et al., AIP Conference Proceedings Vol.~1101, (American Institute of Physics, Melville, NY, 2009), pp.\ 260--279.

\bibitem{Fuchs2013}
C.~A. Fuchs and R.~Schack, ``\href{https://journals.aps.org/rmp/abstract/10.1103/RevModPhys.85.1693}{Quantum-Bayesian Coherence},'' {\sl Reviews of Modern Physics\/} {\bf 85}, 1693--1715 (2013).

\bibitem{Cavalcanti2021}
E. Cavalcanti, ``The View from the Wigner Bubble,'' {\sl Foundations of Physics\/} {\bf 51}, 39 (2021); \href{https://arxiv.org/abs/2008.05100}{\tt arXiv:2008.05100}.

\bibitem{Weber2006}
M.~Weber, {\sl Whitehead's Pancreativism:\ The Basics}, (Ontos Verlag, Frankfurt, 2006).

\bibitem{Weber2011}
M.~Weber, {\sl Whitehead's Pancreativism:\ Jamesian Applications}, (Ontos Verlag, Frankfurt, 2011).

\bibitem{Ulfbeck2001}
O. Ulfbeck and A. Bohr, ``Genuine Fortuitousness. Where Did That Click Come From?''\ {\sl Foundations of Physics\/} {\bf 31}, 757--774 (2001).

\bibitem{Nielsen2020}
M. Nielsen and R. T. Stewart, ``Persistent Disagreement and Polarization in a Bayesian Setting,'' {\sl British Journal for the Philosophy of Science\/} {\bf 72}(1), 51--78 (2020).

\end{thebibliography}
\end{document}